\documentclass[a4paper,11pt]{article}
\usepackage{tablefootnote}
\usepackage[english]{babel}
\usepackage[utf8x]{inputenc}
\usepackage[T1]{fontenc}
\usepackage[flushmargin]{footmisc}
\usepackage{setspace}
\usepackage{apacite}
\usepackage[comma]{natbib}
\usepackage{array}
\usepackage{float}
\usepackage{amsmath}
\usepackage{amsfonts}
\usepackage{amssymb}
\usepackage{ae}
\usepackage{caption}
\usepackage{tkz-graph}
\usepackage{graphicx}
\usepackage{subcaption}
\usepackage{changepage}
\usetikzlibrary{shapes.geometric}

\usepackage[a4paper,left=2.5cm,right=2.5cm,top=3cm,bottom=3cm]{geometry}
\usepackage[colorinlistoftodos]{todonotes}
\usepackage[colorlinks=true, allcolors=blue]{hyperref}
\begin{document} \doublespacing \pagestyle{plain}

\begin{center}

{\LARGE The Impact of Response Measures on COVID-19-Related Hospitalization and Death Rates in Germany and Switzerland}

{\large
	\vspace{1.3cm}}

{\large Martin Huber and  Henrika Langen}\medskip

{\normalsize University of Fribourg, Dept.\ of Economics \bigskip }
\end{center}

\bigskip

\noindent \textbf{Abstract:} {\small We assess the impact of COVID-19 response measures implemented in Germany and Switzerland on cumulative COVID-19-related hospitalization and death rates. Our analysis exploits the fact that the epidemic was more advanced in some regions than in others when certain lockdown measures came into force, based on measuring health outcomes relative to the region-specific start of the epidemic and comparing outcomes across regions with earlier and later start dates. When estimating the effect of the relative timing of measures, we control for regional characteristics and initial epidemic trends by linear regression (Germany and Switzerland), doubly robust estimation (Germany), or synthetic controls (Switzerland).  We find for both countries that a relatively later exposure to the measures entails higher cumulative hospitalization and death rates on region-specific days after the outbreak of the epidemic, suggesting that an earlier imposition of measures is more effective than a later one. For Germany, we also evaluate curfews (as introduced in a subset of states) based on cross-regional variation.  We do not find any effects of curfews on top of the federally imposed contact restriction that banned groups of more than 2 individuals. Finally, an analysis of mobility patterns in Switzerland shows an immediate behavioral effect of the lockdown in terms of reduced mobility.
}

{\small \smallskip }

{\small \noindent \textbf{Keywords: } COVID-19, pandemic, social distancing,  lockdown, treatment effect, synthetic control}

{\small \noindent \textbf{JEL classification: } I18, I12, H12   \quad }

\bigskip
{\small \smallskip {\scriptsize \noindent We have benefited from comments by Cristian Carmeli, Arnaud Chiolero, and Reiner Eichenberger. We are grateful to the Swiss Federal Office of Public Health for providing access to their COVID-19 data, to the COVID-19 platform of the Swiss School of Public Health for their data management, and to Andrea Sommer-Gauch for her excellent research assistance. Addresses for correspondence: Martin Huber, University of Fribourg, Bd.\ de P\'{e}rolles 90, 1700 Fribourg, Switzerland; martin.huber@unifr.ch; Henrika Langen, University of Fribourg, Bd.\ de P\'{e}rolles 90, 1700 Fribourg, Switzerland; henrika.langen@unifr.ch.
}  }

{\thispagestyle{empty}\pagebreak  }

{\small \renewcommand{\thefootnote}{\arabic{footnote}} %
\setcounter{footnote}{0}  \pagebreak \setcounter{footnote}{0} \pagebreak %
\setcounter{page}{1} }

\section{Introduction}\label{intro}

This paper assesses how the COVID-19 response measures implemented in Switzerland and Germany affect the development of cumulative COVID-19-related hospitalization and death rates. In both countries, the federal governments implemented extensive lockdown measures, including the closure of non-essential shops, schools, childcare centers, cafes, bars and restaurants. In Germany, these measures were further enhanced with a ban on gatherings with more than two people decided at federal level and curfews implemented in several states. With the measures in place for some weeks, both countries report a flattening of the COVID-19 epidemic curve. This alone, however, does not necessarily exclusively reflect the impact of the measures, but likely also general time trends in the spread of the virus. For this reason, this study aims to provide evidence about the causal effects of the German and Swiss measures by exploiting variation (i) in their relative timing due the fact that the epidemic was more advanced in some regions than in others when certain measures came into force and (ii) across regions due to the fact that some measures were only introduced in a subset of regions.

A range of studies on the impact of COVID-19 response measures focus on predicting the development of the pandemic in terms of infections, hospitalizations, or death rates based on simulating the spread of the virus and calibrating the model as a function of the measures. For instance, \cite{Singapore} provide a simulation study on the COVID-19 outbreak in Singapore and model the development of COVID-19 infections under four potential intervention scenarios. Likewise, \cite{Bicheretal} developed an agent-based simulation model to predict the development of infections under different scenarios of lockdown timing and exit strategies out of the lockdown in Austria, finding that delaying the lockdown by 1 week would have translated into an increase of infections by 4 times. \cite{DonsimoniApr} simulate the effect of lockdown timing and duration on the rate of COVID-19 infections and the expected end date of the epidemic in Germany. The study suggests that a complete lift of measures on April 20th would have borne the risk of increasing infection rates. The authors further advise to adopt exit strategies and policies that differ across regions in order to learn about which measures are most effective for containing the epidemic while reducing social and economic costs.

In contrast to such simulations, in which empirical data serve for calibrating parameters in prediction models, a growing literature applies policy evaluation methods as outlined in \cite{ImWo08} to assess the effectiveness of lockdown measures based on variation across regions and over time. \cite{Qiu} for instance investigate the influence of socioeconomic factors and COVID-19 response  measures on transmission dynamics in China, finding that measures at a local level have a larger impact on the epidemic curve than restricting population flows between cities. \cite{JuranekZoutman2020} use an event study approach to assess the effect of the lockdown measures of Denmark and Norway on hospitalizations based on a comparison with Sweden whose measures are comparably lenient. Results suggest that the peak number of hospitalizations would have more than doubled in Denmark and Norway had they followed Sweden's strategy. 

\cite{Daveetal2020} use a difference-in-differences approach to evaluate lockdown measures (namely shelter  in  place  orders)  in the US by exploiting variation in responses across states and over time. As a consequence of the measures, they find an important increase (of 5 -10\%) in the rate at which state residents remained in their homes full-time as well as substantial reductions in cumulative COVID-19 cases (44\% after three weeks),\footnote{The estimated effect on fatalities is also negative but less precise.} with early adopting states with a high population density benefitting most. See also \cite{Fowleretal2020} for a related difference-in-differences strategy for the US that suggests reductions in infections, too, as well as in fatalities. Results in \cite{Friedsonetal2020}, who use a synthetic control approach to analyse the measures' effectiveness in California, and \cite{Daveetal2020b}, who evaluate the impact of the measures implemented in Texas in an event study framework, point in the same direction. \cite{Weber2020} exploits regional differences in the timing of measures in Germany finding that school closures, prohibition of mass events, as well as gathering bans and curfews played a major role in reducing the number of confirmed infections, while border closures and shut-downs of the service and retail sector did not show a significant effect. Studies on the impact of face mask requirements in public transport, retailers and public businesses find evidence for a reduction in the spread of the virus through such requirements, see e.g. \cite{Mitze2020} for a synthetic control study on German data and \cite{Chernozhukov2020}, who assess the impact of such requirements in the US within a causal framework that allows for both, direct effects of COVID-19 response measures and indirect effects through behavioral changes.

\cite{AskitasTatsiramosVerheyden2020} apply an event study design to assess a range of different response measures across 135 countries and find that cancelling public events and restricting gatherings reduce new infections more effectively than mobility restrictions like international travel controls. This is in line with \cite{Bonardietal2020} who consider first difference and AR(1) models based on 184 countries and conclude that lockdown measures generally reduce confirmed infections and fatalities (and even more so if imposed rather earlier than later), while border closures do not show important effects. Findings in \cite{Banholzer2020}, a study on 20 Western countries in a Bayesian framework, suggest that venue closures and gathering bans are most effective in reducing infections but also attest a significant effect of border closures. 

Our paper contributes to this growing literature by analysing COVID-19-related hospitalizations and death rates across administrative units over time, namely across counties in the case of Germany and across cantons in the case of Switzerland. We estimate the effect of the relative timing of lockdown measures based on measuring health outcomes relative to the region-specific start of the epidemic and comparing outcomes across regions with earlier and later start dates. The start date is defined as the day on which the confirmed regional infections per 10,000 inhabitants exceed 1 for the first time. In the analysis, we control for regional characteristics (population size and density, age structure, and GDP per capita), initial trends of the epidemic (median age of confirmed infections and initial growth rate of confirmed infections), and other policies selectively introduced prior to the major lockdowns (e.g.\ a ban on visits to hospitals and retirement homes in some regions).

Linear regression estimates suggest that for both Switzerland (which also includes the Principality of Liechtenstein as data point) and Germany, a relatively later exposure to the measures entails higher cumulative hospitalization and death rates on sufficiently advanced region-specific days after the outbreak of the epidemic. This suggests that an earlier imposition of measures is more effective than a later one w.r.t.\ our health outcomes, which is in line with findings in \cite{amuedo2020timing} on the effect of lockdown timing on COVID-19-related deaths in Spain. For Germany with its substantially larger number of observations, we also estimate the effect of the relative timing based on doubly robust (DR) estimation, see \cite{Robins+94} and \cite{RoRo95}, which is a more flexible approach than exclusively relying on a linear outcome model. For Switzerland, we also consider the synthetic control method, see \cite{Abadie03} and \cite{Abadieetal2010}, to assess for two selected cantons with a relatively late exposure what their counterfactual outcomes would have been under an earlier exposure. Both the DR and synthetic control methods corroborate the findings of the linear regression. For Germany only, we also evaluate the effect of curfews that were introduced by a subset of German states in addition to the federal lockdown measures and bans of gatherings with more than two individuals. Exploiting this cross-sectional variation while controlling for observed characteristics, neither linear regression nor DR estimation suggest that curfews further reduce hospitalizations and fatalities under the lockdown measures already in place, which is in line with the findings in \cite{Bonardietal2020} and \cite{Banholzer2020}. Finally, we investigate how mobility patterns in Switzerland changed after the lockdown measures by means of a t-test that is applied to canton-specific mobility statistics and find an immediate behavioral change in terms of reduced mobility. This is particularly interesting in light of the results in \cite{Debetal2020}, a cross-country study on Apple mobility trends, which suggest that social distancing measures have been more effective in countries where they resulted in larger mobility reductions.

The remainder of this paper is organized as follows. Section \ref{measures} provides an overview of the timeline of COVID-19 measures in Switzerland and Germany. Sections \ref{data} and \ref{methods} describe the data and econometric methods used in the analyses. Section \ref{results} presents and interprets the results. Section \ref{conclusion} concludes.

\section{Timeline of COVID-19 Response Measures}\label{measures}
Both Germany and Switzerland are federal states with competencies in epidemic control partly belonging to the 26 cantons in Switzerland and the 16 federal states (Länder) in Germany. The German states themselves are comprised of all in all 401 counties (Kreise) which also have certain competencies in handling epidemic outbreaks. With competencies fragmented across the federal governments and sub-federal authorities, not all measures were implemented in all regions and, if so, not always at the same time. However, decisions on key COVID-19 response measures were made at the federal level in both countries.

In Switzerland, the first COVID-19 response measure, a ban of events with more than 1000 visitors, was announced and implemented at the federal level on February 28th when there were some 25 confirmed COVID-19 cases (0.03 per 10,000 inhabitants) in Switzerland. Several measures at the cantonal level followed. For instance, many cantons introduced a ban on visits to retirement homes. Some 2.5 weeks after the first measure was implemented, the Federal Council decided to close all schools and childcare centers in Switzerland as well as non-essential shops, cafes, bars, and restaurants on March 16th. In the following, we will refer to these measures as lockdown measures. At that point in time, the rate of confirmed infections in Switzerland was at 4.2 per 10,000 inhabitants. The schedule of response measures in the Principality of Liechtenstein (LI) was similar to that in Switzerland with the lockdown entering into force two days later. Due to the two countries' similar schedules of COVID-19 response measures, their geographic proximity and their economic, cultural and political interconnection, we include LI as additional data point when investigating the impact of the lockdown measures in Switzerland.

In Germany, first measures at the federal level were implemented between March 9th and March 12th. On March 8th, when there were some 1000 reported COVID-19 cases (0.12 per 10,000 inhabitants) in Germany, the federal government advised against events with more than 1000 visitors. This recommendation was translated into a ban by most federal states, while others implemented it as recommendation only. As in Switzerland, schools and childcare centers in most German states closed on March 16th, the remaining states followed within two days. The closure of all non-essential retailers, bars and public events of any kind and the restriction of restaurant opening hours was decided at the federal level on March 16th when the overall rate of confirmed infections reached 1.1 per 10,000 inhabitants. The states implemented these measures between March 17th and March 20th. Other than in Switzerland and LI, these measures were further enhanced later on. On March 22nd, a ban of groups with more than two individuals was decided at the federal level and several states additionally implemented curfews. Since April 17th, more and more states have made wearing face masks in shops and public transport compulsory, resulting in a nationwide requirement to wear masks in public from April 27th on. Meanwhile, lockdown measures have been lifted gradually in Switzerland and Germany, with distinct schedules and exit strategies across countries and states. For instance, curfews ended in the respective German states on April 27th, with the exception of Bavaria, where they ended on May 5th.

\section{Data}\label{data}

For Switzerland and LI, data on  confirmed COVID-19 infections as well as on COVID-19-related hospitalizations and deaths are amalgamated by the Swiss Federal Office of Public Health (FOPH) and made available to the interuniversity research consortium of the Swiss School of Public Health (www.ssphplus.ch). For each confirmed case, the FOPH gathers information on the reporting canton, test date, as well as patient's age and gender from laboratory declarations. %which is complemented with additional information on hospitalized patients stemming from clinical declarations.
%The data are provided in long format with each row representing one confirmed COVID-19 case.
For our analysis, we aggregate the number of confirmed infections, hospitalizations and fatalities by canton and test date, compute the respective cumulative numbers by canton and date, and complement the data with socio-demographic variables at the cantonal level (and for LI) from the statistical offices of Switzerland and LI. For each of the 26 Swiss cantons and LI, we calculate the rate of cumulative confirmed infections, hospitalizations and fatalities per 10,000 inhabitants, as well as the median age of those tested positively for COVID-19 prior to the lockdown measures in Switzerland and LI. Furthermore, we construct indicators for whether a canton has introduced certain additional measures not imposed by the federal government along with variables providing the start date of such canton-level measures as stated in press releases of the respective cantons.
%Additionally, we analyze data on the weekly numbers of registered deaths since 2015 published by the Swiss Federal Office of Statistics with the numbers for 2020 being provisional and including COVID-19-related deaths.
Additionally, we investigate data on mobility patterns of a representative sample of the Swiss adult population before and after the implementation of the lockdown measures. The data are collected with a GPS tracking app in the MOBIS-COVID19 study, a research project initiated by the ETH Zurich and the University of Basel, see \cite{Molloyetal2020}. The participants' mobility patterns have been tracked since autumn 2019 allowing for comparisons between mobility patterns before and after the implementation of main response measures in Switzerland.

In Germany, all confirmed infections and deaths are reported to the Robert Koch Institute (RKI), a federal government agency and research institute for disease control and prevention. The RKI publishes data on the age group, gender, test date and county of residence of each validated COVID-19 case reported to the institute. Only for the county of Berlin with 3.6 million inhabitants, the RKI also reports the urban residential district of confirmed cases. All in all, there are 401 counties in Germany and 12 residential districts in Berlin. Similar to Switzerland, we aggregate the data by county (or residential district, respectively) and test date, and compute cumulative confirmed cases and fatalities by county and date. We complement the data with socio-demographic variables at the county/district level from the Federal Office of Statistics, the statistical offices of the federal states and the statistical office of the city of Berlin. As most measures in Germany were implemented at the state or even county level and at different points in time, we generate variables for all measures indicating whether and when they were imposed in each county.

\begin{figure}[!h]
	\centering
	\begin{subfigure}[b]{0.49\textwidth}
		\includegraphics[width=\textwidth]{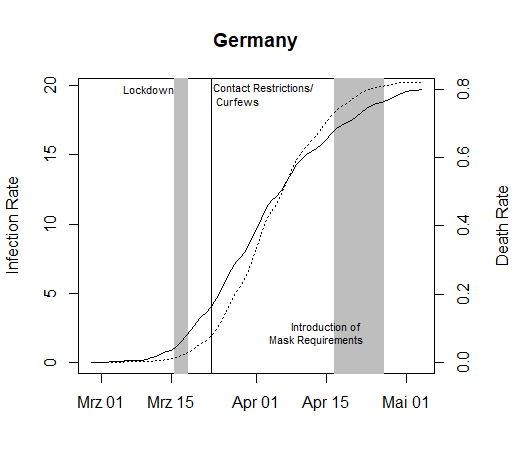}
	\end{subfigure}
	\begin{subfigure}[b]{0.02\textwidth}
	\end{subfigure}
	\begin{subfigure}[b]{0.49\textwidth}
		\includegraphics[width=\textwidth]{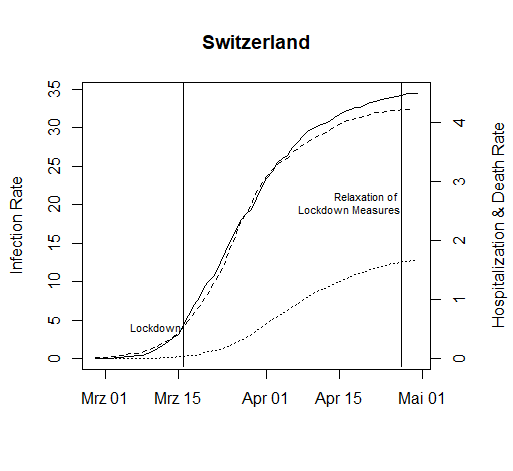}
	\end{subfigure}
	\caption{{\small \textit{Cumulative confirmed infections (solid line), deaths (dotted line) and hospitalizations (dashed line) per 10,000 inhabitants in Germany and Switzerland.}}}\label{fig:descriptives}
\end{figure}

\begin{figure}[!h]
	\centering
	\begin{subfigure}[b]{0.49\textwidth}
		\includegraphics[width=\textwidth]{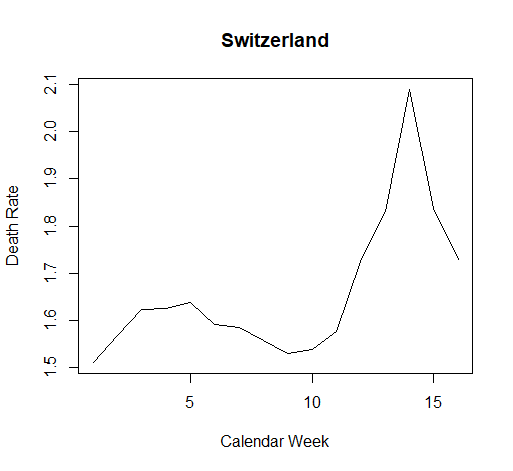}
	\end{subfigure}
	\begin{subfigure}[b]{0.49\textwidth}
		\includegraphics[width=\textwidth]{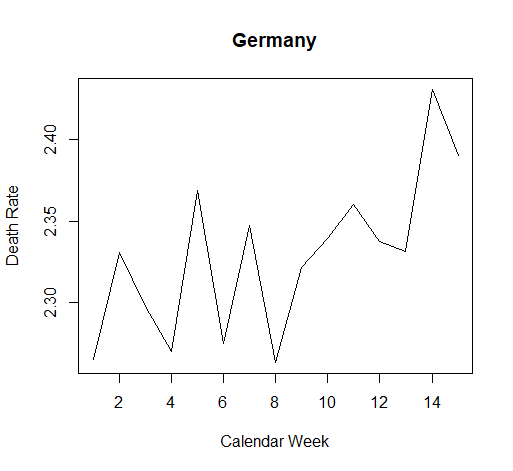}
	\end{subfigure}
	\caption{{\small \textit{Overall deaths per 10,000 inhabitants by calendar week in Switzerland (left) and Germany (right). Source: federal statistical offices of Switzerland (\href{https://www.bfs.admin.ch/bfs/en/home/statistics/population/births-deaths/deaths.html}{www.bfs.admin.ch}) and Germany (\href{https://www.destatis.de/EN/Themes/Society-Environment/Population/Deaths-Life-Expectancy/_node.html}{www.destatis.de}), retrieval date: May 6th.}}}\label{fig:descriptivedeathrates}
\end{figure}

Figure \ref{fig:descriptives} provides the cumulative numbers of confirmed COVID-19 infections and COVID-19-related deaths per 10,000 inhabitants in Germany (left) as well as cumulative numbers of confirmed infections, hospitalizations and deaths in Switzerland (right). The figure suggests a flattening of the COVID-19 epidemic curve in both countries after the main COVID-19 measures have been in place for some weeks, which does, however, not necessarily exclusively reflect the causal impact of the measures. As a further descriptive statistic, Figure \ref{fig:descriptivedeathrates} provides the overall deaths per 10,000 inhabitants (thus including COVID-19-related mortality) by calendar week in Germany and Switzerland since January  1st 2020 (provisional data). While the increase in mortality in March and April can be linked to the COVID-19 epidemic (a finding that also holds when controlling for the average mortality over 2015-2019), we cannot directly infer how large the increase would have been with and without the lockdown measures. For this reason, our analysis aims at shedding light on the causal effect of the measures.

%\begin{figure}[!h]
%	\centering
%	\begin{subfigure}[b]{0.49\textwidth}
%		\includegraphics[width=\textwidth]{Figures/deathrate.png}
%	\end{subfigure}
%	\begin{subfigure}[b]{0.49\textwidth}
%		\includegraphics[width=\textwidth]{Figures/Ddeathrate.png}
%	\end{subfigure}
%	\caption{{\small \textit{Overall rate of deaths per 10,000 inhabitants in Switzerland (left) and Germany (right).}}}
%\end{figure}

\section{Econometric Approach}\label{methods}

In our analysis, we exploit the fact that the epidemic was more advanced in some regions than in others when the key control measures came into force. In Switzerland, for instance, Basel-Stadt had already more than 1 confirmed case per 10,000 inhabitants 12 days before the federal lockdown measures were implemented, while other cantons such as St.\ Gallen were at an earlier stage, reaching 1 confirmed infection per 10,000 inhabitants on the day of the lockdown. In Germany, the county of Heinsberg recorded more than 1 confirmed infection per 10,000 inhabitants already 19 days before the lockdown. In several other counties this level of infections was reached only after the lockdown.

For Germany, we investigate the impact of the lockdown measures as well as the curfew on cumulative deaths per 10,000 inhabitants. For Switzerland and LI, we assess the causal effect of the lockdown on both cumulative hospitalizations and deaths per 10,000 inhabitants. The idea is to quantify the epidemic stage of each canton/county when measures were implemented by defining dates on which the health outcomes are measured relative to the day a canton/county first reached a certain rate of confirmed infections. For both Germany and Switzerland, we define the start date of the epidemic as the day when the rate of infections first reached or exceeded 1 infection per 10,000 inhabitants. In Switzerland, for instance, the start date of the epidemic in Basel-Stadt is on March 5th (late exposure to measures) while in St.\ Gallen the epidemic started on March 16th (early exposure to measures). Appendix \ref{app:Startdates} provides the start states for all Swiss cantons and LI.

Besides their obvious relevance for health care, a further motivation to consider hospitalization and death rates as outcomes is that their measurement is likely more robust to differences in testing strategies across regions than the measurement of confirmed COVID-19 infections. While the share of infections with mild symptoms being detected ceteris paribus likely rises with increased testing, the number of hospitalizations and fatalities gives a better estimate of the severeness of the epidemic in terms of human loss and strains for the health care system. %Of course, hospital policies on COVID-19 testing have changed  during the very early stage of the epidemic outbreak in Europe but we concentrate our analysis on the time when COVID-19 testing has already become standard clinical practice.
As both Germany and Switzerland maintain a system of mandatory health insurance and neither country generally saw their hospitalization capacities exhausted, we would suspect that the number of COVID-19-related hospitalizations in general mirrors well the number of individuals infected with COVID-19 that are in need of hospitalization. Nevertheless, a potential concern in our analysis is that the criteria for hospitalizations might not be uniform across regions. The same may apply to the measurement of fatalities, i.e.\ the definition of criteria according to which a decease is attributed to COVID-19. If such measurement issues in health outcomes are not systematically associated with the region-specific start date of the epidemic (or more generally, with the policy interventions considered), they do not bias the results of our analysis. However, if for instance regions with an earlier start date and a more advanced epidemic systematically applied more stringent rules for hospital admissions (e.g.\ to prevent capacity constraints), this could also entail an underestimation of COVID-19 fatalities due to underreporting deceases at home. In this case, our analysis of the relative timing of measures presented below would likely provide a lower bound of the true effect on (capacity-unconstraint) hospitalizations and fatalities.

%Further, the number of deaths should not include deaths due to lack of hospital capacities.

\subsection{OLS Approach} \label{OLS}

We compare the average number of cumulative hospitalizations and fatalities per 10,000 inhabitants on canton/county-specific epidemic days across three groups of cantons/counties. These groups are defined by the canton/county-specific epidemic day when lockdown measures came into place. For Switzerland and LI, we distinguish the groups of cantons as follows. Cantons reaching or exceeding 1 confirmed infection per 10,000 inhabitants at most 4 days before the lockdown measures are exposed to the measures at a relatively early stage of the epidemic and constitute the reference group (sample size $N = 8$). Those cantons with at least 1 confirmed infection per 10,000 inhabitants between 5 and 8 days before March 16th (or March 18th in the case of LI) are the intermediate intervention group ($N = 11$). Those with a canton-specific start date at least 9 days before March 16th  are the late intervention group ($N = 8$).

For Germany, we proceed analogously and define the treatment groups based on the days between the county-specific start of the epidemic and the lockdown according to the retail closures between March 17th and 20th, but with somewhat different time brackets. Counties with at least 1 confirmed infection per 10,000 inhabitants not earlier than 3 days after the implementation of lockdown measures make up the reference group. The specified start dates are later than the lockdown, which may at first glance raise endogeneity concerns. However, any effect of the measures can materialize in the outcomes only with a substantial time lag of more than 1.5 weeks (due to incubation time and reporting lags), as also confirmed in our analysis. Therefore, confirmed infection rates are not yet influenced by the measures even several days after the lockdown. Yet, we exclude 4 counties having start dates as late as 9 days after the lockdown or later, leaving us with a reference group of $N = 52$. The intermediate intervention group is comprised of all counties with at least 1 confirmed infection per 10,000 inhabitants between 3 days before and 2 days after the lockdown ($N = 275$). The late intervention group consists of counties with at least 1 confirmed infection per 10,000 inhabitants more than 3 days before the lockdown ($N = 81$).

We estimate the difference in cumulative death rates, as well as hospitalization rates for Switzerland and LI, between either of the two treatment groups (intermediate and late intervention group) and the reference group by means of an OLS regression with treatment indicators. We also control for the following canton-/county-specific covariates: population size and density, income per capita, age distribution, age structure of positively tested up to the lockdown, the initial canton-/county-specific growth trend for confirmed cases, and canton-specific bans on visits in hospitals and retired homes entering into force prior to the lockdown. The large number of counties in Germany allows us to further control for past mortality by age group, past mortality rate related to respiratory diseases and hospital capacities (beds/1000 inhabitants). We also control for state-specific measures entering into force prior to the general lockdown, like bans of or recommendations against events with more than 1000 visitors, as well as curfews imposed in some states only a few days after the general lockdown. Appendix \ref{app:Descriptives} provides descriptive statistics of the covariates used in the analysis of the German and Swiss measures for the respective total samples as well as separately for the various intervention groups.

Though aiming to control for confounders jointly affecting the region-specific epidemic and the health outcomes in a comprehensive way, we cannot completely rule out that some important characteristics are omitted in our analysis. For instance, we cannot directly control for the amount of inter-generational interactions, which is according to \cite{Bayer} correlated with the ratio of deaths over confirmed cases and could potentially differ across regions. We, however, point out that the results for the relative timing of measures are quite robust to (not) controlling for covariates. Since the lockdown measures in Germany and in Switzerland have been eased starting with April 20th and April 27th, respectively, we evaluate the effect of the relative timing of measures on the health outcomes in these countries until April 23rd and April 30th, respectively.

For Germany, we also investigate the impact of curfews, as introduced in some federal states between March 21st and 23rd on top of the federally imposed contact restriction that banned groups of more than 2 individuals. The OLS regression contains a binary treatment indicator for curfews as well as a range of control variables. The latter include the previously mentioned county-specific characteristics, growth trends and COVID-19 response measures, and in addition the cumulative confirmed infections and death rates on several days prior to the curfews, in order to make regions exposed and not exposed to curfews as similar as possible. The OLS specification is provided in Appendix \ref{app:OLSEstimates}, descriptive statistics for counties with and without curfews in Appendix \ref{app:Descriptives}.

\subsection{Doubly Robust Estimation} \label{semiparametric}

The larger number of regions in Germany allows us to also consider a more flexible (so-called semiparametric) evaluation approach based on doubly robust (DR) estimation, see \cite{Robins+94} and \cite{RoRo95}. It is based on (i) estimating a logit model for the treatment probability as a function of the covariates as well as a linear model for the outcome as a function of the treatment and the covariates and (ii) using the respective model predictions as plug-in parameters for the estimation of the treatment effects. DR provides consistent effect estimates if at least one of the plug-in models is correctly specified and thus relies on less stringent assumptions than OLS. Using the `drgee' package of \cite{DoublyRobustEstimationwiththeRPackagedrgee} for the statistical software `R', we apply DR for estimating the average effect of a binary intervention separately to subsets of counties consisting of the reference group and either the intermediate intervention group or the late intervention group.

%As described in the last section, we distinguish three groups of counties based on the county-specific epidemic day when lockdown measures came into place. In the OLS approach, we control for all county-specific characteristics that potentially confound the association between the treatment (belonging to \textit{intermediate} or \textit{late intervention group} respectively) and the outcome (cumulative death rate). For this to yield unbiased results the model must be correctly specified, i.e. no potential confounder must be left out and the association must be linear in parameters.

%The idea of the DR semiparametric approach is to estimate two separate models, called nuisance models: one logit model for the association between treatment exposure and potential confounders, and one linear model for the association between the outcome and potential confounders. The estimate for the treatment-outcome association resulting from combining both nuisance models is unbiased as long as one of the two nuisance models is correctly specified.

\subsection{Synthetic Control Approach} \label{synthetic}

For Switzerland, we complement the regression analysis with a synthetic control approach, a quantitative case study method suggested in \cite{Abadie03}. To this end, we compare cumulative hospitalization and fatality rates in a specific canton with a late exposure to the lockdown to the rates of an artificially (or synthetically) created counterfactual canton. This synthetic canton should be comparable to the original reference canton in terms of covariates outlined in Section \ref{OLS} and pre-treatment health outcomes (measured 2 and 5 days after the start date), but characterized by an earlier exposure to the lockdown.\footnote{In contrast to the OLS specification provided in Appendix \ref{app:OLSEstimates}, squared variables (i.e.\ the squares of the population share aged 65+ and of the median age of confirmed infections prior to the lockdown) are not included. In addition, the dummy for the number of inhabitants being smaller than 60,000 is replaced by the actual number of inhabitants.} To this end, the synthetic canton is generated as a weighted average of control cantons with an earlier exposure using the `Synth' package of \cite{Abadieetal2011} for the statistical software `R', where the weights depend on how close their characteristics and pre-treatment outcomes match the values of the reference canton with the later exposure. The control pool includes all in all 11 cantons that reached 1 confirmed infection per 10,000 inhabitants at most 3 days before the lockdown.

\section{Results}\label{results}
\subsection{Germany}
%\subsubsection{OLS Approach} \label{DOLSResults}

\begin{figure}[!h]
	\centering
	\begin{subfigure}[b]{0.49\textwidth}
		\includegraphics[width=\textwidth]{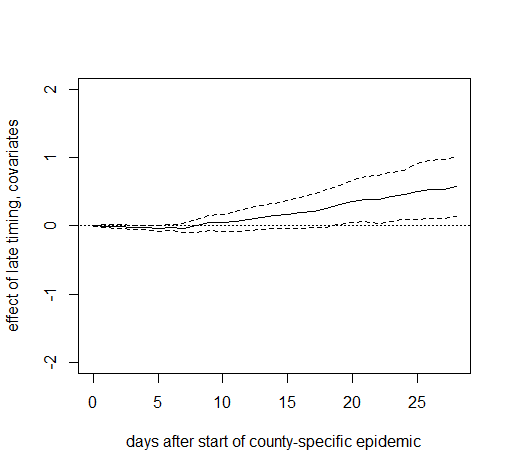}
	\end{subfigure}
	\begin{subfigure}[b]{0.49\textwidth}
		\includegraphics[width=\textwidth]{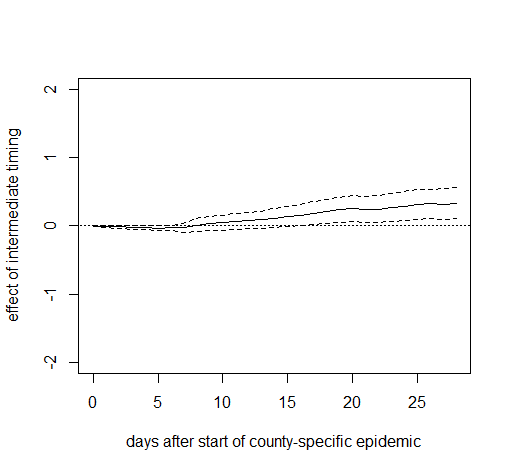}
	\end{subfigure}
	\caption{{\small \textit{OLS effects of late (left) and intermediate (right) timing of measures on cumulative deaths per 10,000 inhabitants in Germany.}}}\label{fig:DOLS}
\end{figure}

Figure \ref{fig:DOLS} reports the mean differences in cumulative fatalities per 10,000 inhabitants between either treatment group and the early intervention group (reference group) per day up to 28 days after the county-specific start date (solid lines) based on the OLS approach.\footnote{The motivation for the 28 days window is that we would like to include all (but 4) counties while at the same time only considering the period when the lockdown measures were fully implemented. As the last county we include in our evaluation sample saw its start of the epidemic 8 days after the lockdown, the time range considered in the analysis is limited to this specific window not including any effects of the first easing of lockdown measures starting with April 20th.} It also includes 90\% confidence intervals (dashed lines). The mean differences in fatality rates between the late and the early intervention groups (left) remain close to zero during the first 2.5 weeks of the county-specific epidemic but show a positive and statistically significant tendency thereafter. The point estimates suggest that  after one month, fatalities per 10,000 inhabitants are reduced by 0.7 cases under an earlier lockdown. Also the difference in death rates between the intermediate and the early intervention groups are statistically significant at the 10 percent level, but (expectedly) smaller in magnitude. Overall, the results suggest that the relative timing of measures had a perceptible impact on COVID19-related fatalities in Germany. We note that Appendix \ref{app:OLSEstimates} provides the OLS specification with the full list of coefficients on treatments and covariates along with standard errors 28 days after the start of the epidemic. Concerning the robustness of our findings, we note that estimations without controlling for observed covariates  yield qualitatively similar results, see Appendix \ref{app:DOLS}). %as well as with different definitions of the start dates(0.4 confirmed infections per 10,000 inhabitants) and treatment groups

\begin{figure}[htp]
	\centering
	\begin{subfigure}[b]{0.49\textwidth}
		\includegraphics[width=\textwidth]{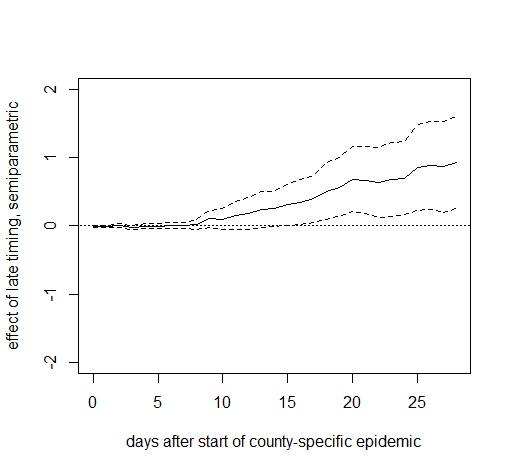}
	\end{subfigure}
	\begin{subfigure}[b]{0.49\textwidth}
		\includegraphics[width=\textwidth]{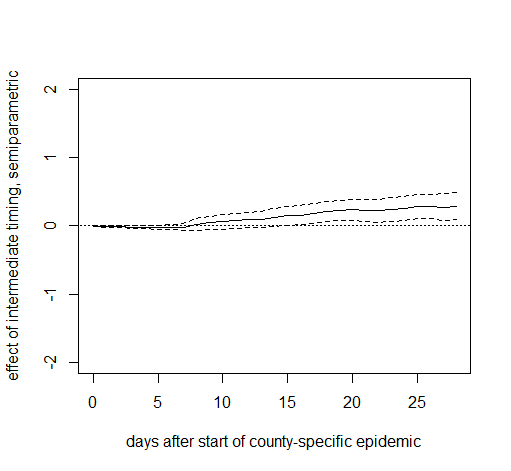}
	\end{subfigure}
	\caption{{\small \textit{DR effects of late (left) and intermediate (right) timing of measures on cumulative deaths per 10,000 inhabitants in Germany.}}}\label{fig:Dsemi}
\end{figure}

Figure \ref{fig:Dsemi} reports the estimates of DR, which are generally similar to OLS, though suggesting an even stronger effect of a late timing of lockdown measures on the death rate. The point estimate suggests that an earlier lockdown reduces fatalities by roughly 1 case per 10,000 one month after the start of the epidemic. %Estimations with different  definitions of the start dates yield similar results, see Appendix \ref{app:DOLS}.

\begin{figure}[!h]
	\centering
	\begin{subfigure}[b]{0.25\textwidth}
	\end{subfigure}
	\begin{subfigure}[b]{0.49\textwidth}
		\includegraphics[width=\textwidth]{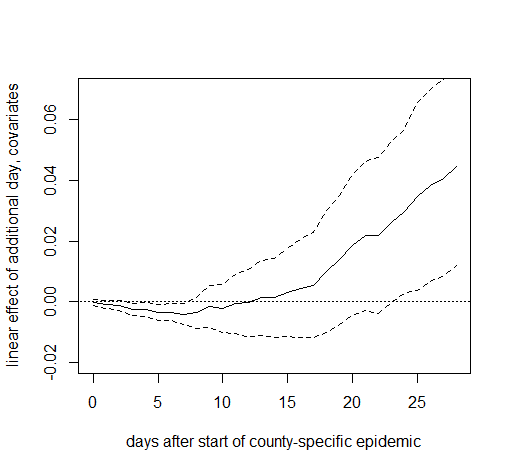}
	\end{subfigure}
	\begin{subfigure}[b]{0.25\textwidth}
	\end{subfigure}
	\caption{{\small \textit{OLS effect of delaying lockdown by one day on deaths per 10,000 inhabitants in Germany.}}}\label{fig:addday}
\end{figure}

Figure \ref{fig:addday} reports the results of a further OLS regression, in which the treatment indicators for the intermediate and late intervention groups are replaced by the time lag between the county-specific start date of the epidemic and the lockdown, in order to (linearly) estimate the effect of the lag. This can be interpreted as the average effect of waiting an additional day before implementing the measures. The point estimates suggest that each additional day without lockdown entails on average 0.04 to 0.05 additional fatalities per 10,000 inhabitants after one month of the epidemic, even though the confidence intervals are rather wide (but yet do not include a zero effect). Again, these results are quite robust to not controlling for covariates, see Appendix \ref{app:DOLS}. %and a different definition of county-specific start dates of the epidemic

\begin{figure}[!h]
\centering
	\begin{subfigure}[b]{0.49\textwidth}
		\includegraphics[width=\textwidth]{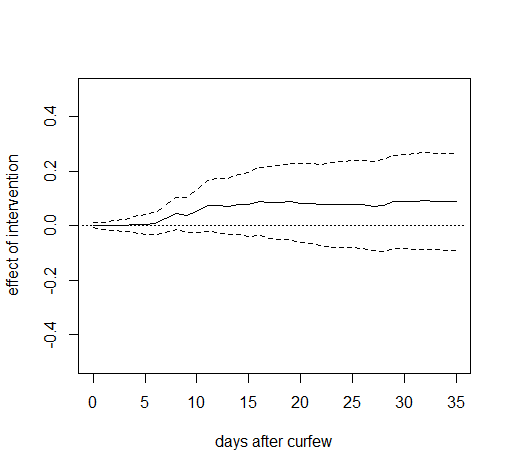}
	\end{subfigure}
	\begin{subfigure}[b]{0.49\textwidth}
		\includegraphics[width=\textwidth]{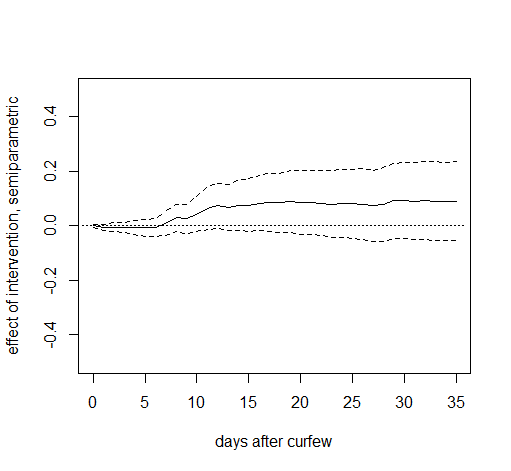}
	\end{subfigure}
	\caption{{\small \textit{OLS (left) and DR (right) effects of curfews on deaths per 10,000 inhabitants in Germany.}}}\label{fig:DAusgangssperre}
\end{figure}

Furthermore, the left graph in Figure \ref{fig:DAusgangssperre} provides the OLS-based effects of curfews relative to contact restrictions, i.e.\ bans of gatherings with more than 2 persons, under all other lockdown measures already in place. The estimates have a positive sign, which appears counter-intuitive as curfews are more restrictive than contact restrictions, but are never statistically significantly different from zero throughout the evaluation window which starts on March 23rd and ends 35 days later. The same finding applies to estimation results based on DR, which are shown in the right graph of Figure \ref{fig:DAusgangssperre}. Therefore, we do not find evidence that curfews are more effective than banning groups for reducing fatality rates.

\subsection{Switzerland and LI}
%\subsubsection{OLS Approach} \label{CHOLSResults}

\begin{figure}[htp]
	\centering
	\begin{subfigure}[b]{0.49\textwidth}
		\includegraphics[width=\textwidth]{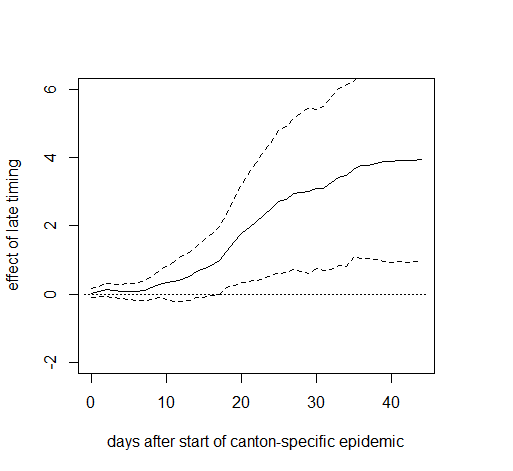}
	\end{subfigure}
	\begin{subfigure}[b]{0.49\textwidth}
		\includegraphics[width=\textwidth]{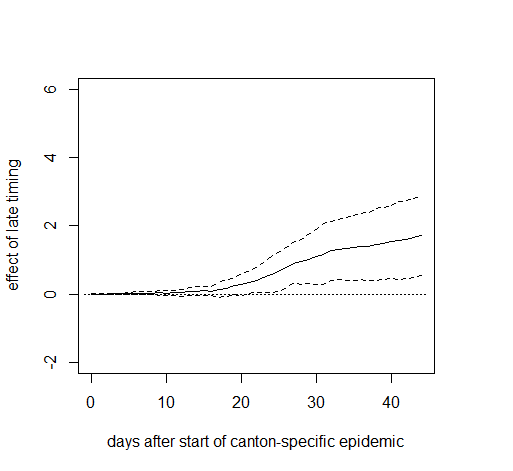}
	\end{subfigure}
	\caption{{\small \textit{Effect of late timing of measures on cumulative hospitalizations (left) and deaths (right) per 10,000 inhabitants.}}}\label{fig:latetiming}
\end{figure}

Figure \ref{fig:latetiming} reports the OLS estimates of the mean differences in cumulative hospitalizations (left) and fatalities (right) per 10,000 inhabitants between the late and the early intervention groups up to 44 days after the start of the canton-specific epidemic (solid line), as well as 90\% confidence intervals (dashed lines). See Appendix \ref{app:OLSEstimates} for the full OLS specification with the coefficients on treatments and covariates on the last day of the evaluation window and fatalities as outcome variable. %As the reference group in the Swiss data is comprised of cantons with canton-specific day 0 on the date of lockdown implementation or maximum 3 days earlier, the figures in this section do not only indicate canton-specific epidemic day on which the lockdown showed a significant effect. They also give an (approximate) idea of the time after lockdown implementation when they first showed a significant effect on COVID-19-related death and hospitalization rates.
We note that the canton of Ticino is excluded from this analysis due to its comparably strong economic and social ties with Northern Italy (which was particularly severely affected by the COVID19 crisis), as this could arguably have affected the canton's hospitalizations and fatalities. However, our findings are quite similar when including Ticino in the regression, as well as when not controlling for covariates, see Appendix \ref{app:OLS}.

\begin{figure}[htp]
	\centering
	\begin{subfigure}[b]{0.49\textwidth}
		\includegraphics[width=\textwidth]{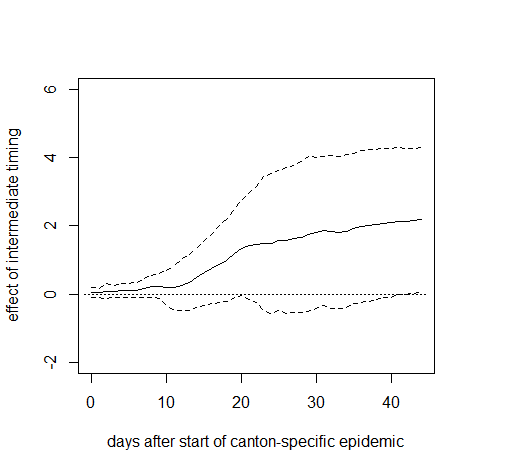}
	\end{subfigure}
	\begin{subfigure}[b]{0.49\textwidth}
		\includegraphics[width=\textwidth]{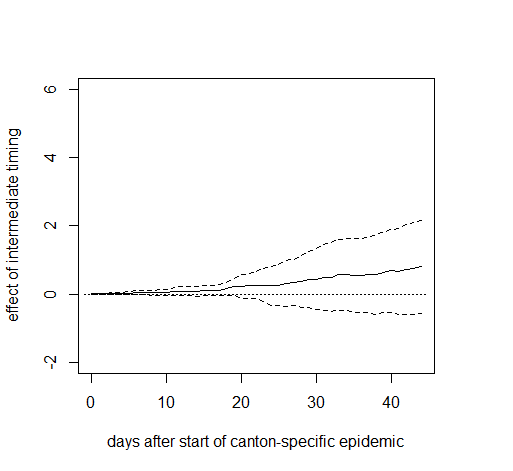}
	\end{subfigure}
	\caption{{\small \textit{Effect of intermediate timing of measures on cumulative hospitalizations (left) and deaths (right) per 10,000 inhabitants.}}}\label{fig:intertiming}
\end{figure}

As for Germany, we see no immediate effect of the relative timing of measures on the health outcomes right after their introduction. However, after about two weeks, there is a positive tendency in the effect on cumulative hospitalizations that becomes statistically significant at the 10\% level about 2.5 weeks after the start of the canton-specific epidemic. The point estimates suggest that after 1.5 months, cumulative hospitalizations per 10,000 inhabitants increase by almost 4 cases when introducing
the measures later rather than earlier, even though the estimates are not very precise (i.e.\ confidence intervals are wide). A qualitatively similar pattern is observed for the effect on cumulative deaths, which becomes statistically significant after about 3 weeks. The point estimates suggest an increase of 1 to 2 fatalities per 10,000 inhabitants in the case of a later lockdown, but precision is again low. Figure \ref{fig:intertiming} reports the same analysis for a comparison of the groups with intermediate and early timing. As these two groups are more similar in terms of the relative timing of the measures, differences are less pronounced and never statistically significant in all but one case, which might be due to low statistical power related to the small number of cantons.\footnote{For cumulative fatalities, we also run the OLS regression using an alternative data source based on calculations of the statistics office of the canton of Zurich, available at \href{https://statistik.zh.ch/internet/justiz_inneres/statistik/de/covid19.html}{https://statistik.zh.ch} (retrieved on May 15th). We obtain a comparable pattern.  Namely, the  late intervention effect turns statistically significant after about 3 weeks with even somewhat higher point estimates (approaching 3) at the end of the evaluation window. The intermediate intervention effect is again insignificant.}

\begin{figure}[htp]
	\centering
	\begin{subfigure}[b]{0.49\textwidth}
		\includegraphics[width=\textwidth]{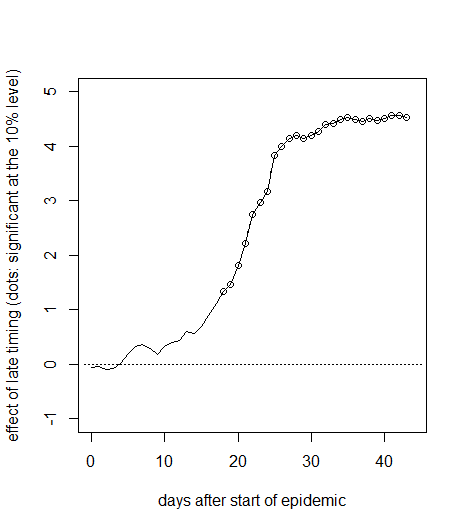}
	\end{subfigure}
	\begin{subfigure}[b]{0.49\textwidth}
		\includegraphics[width=\textwidth]{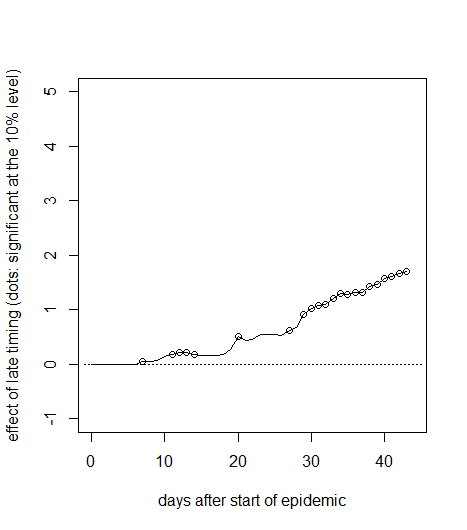}
	\end{subfigure}
	\caption{{\small \textit{Effect of late timing of measures on cumulative hospitalizations (left) and deaths (right) per 10,000 inhabitants in Basel-Stadt.}}}\label{fig:BStiming}
\end{figure}

Finally, we report the results of the synthetic control method for two cantons experiencing the lockdown rather late relative to their start date of the epidemic. Figure \ref{fig:BStiming} plots the difference in cumulative hospitalizations (left) and deaths (right) per 10,000 inhabitants on a daily base after the canton-specific start date  between Basel-Stadt, which was on day 12 of the epidemic when the measures came into force, and its synthetic counterfactual. The latter is generated from a control group of 11 cantons with an earlier timing (with start dates between 3 days before and 1 day after the lockdown). Dots on the solid line imply that the differences are statistically significant at the 10\% level according to placebo tests in the control group, in which each of the 11 cantons is considered as (pseudo-)treated in a rotating scheme in order to estimate its (pseudo-)counterfactual based on the remaining 10 cantons. We, however, note that the estimation of p-values might be imprecise, due to the low number of control cantons available for the placebo tests.

Again, the relative timing of measures shows no immediate effect on hospitalizations but the difference becomes statistically significant after roughly 2.5 weeks. The point estimates suggest that the hospitalization rate in Basel-Stadt could have been reduced by more than 4 hospitalizations if the lockdown measures had been introduced earlier. Similarily, the fatalities per 10,000 inhabitants could have been reduced by 1 to 2 cases about 1.5 months after the start of the epidemic. As for the OLS analysis, the exact numbers should, however, be interpreted with caution, as they are imprecisely estimated and canton-specific factors not considered in the analysis could play a role as well.  %Nevertheless, the analysis suggests that the measures are effective for reducing hospitalizations and that their timing matters.

\begin{figure}[htp]
	\centering
	\begin{subfigure}[b]{0.49\textwidth}
		\includegraphics[width=\textwidth]{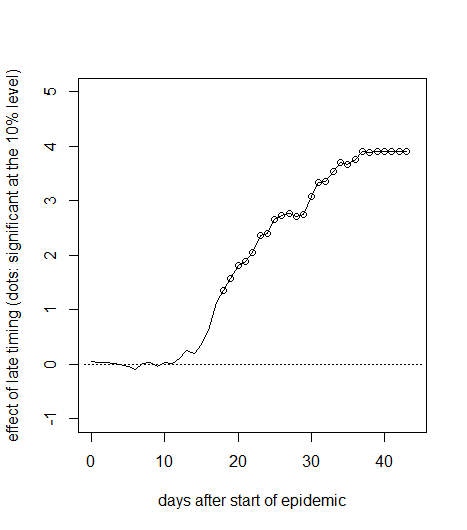}
	\end{subfigure}
	\begin{subfigure}[b]{0.49\textwidth}
		\includegraphics[width=\textwidth]{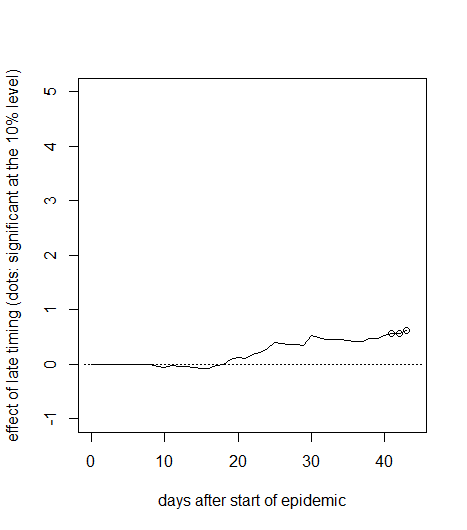}
	\end{subfigure}
	\caption{{\small \textit{Effect of late timing of measures on cumulative hospitalizations (left) and deaths (right) per 10,000 inhabitants in Neuchâtel}}}\label{fig:NEtiming}
\end{figure}

Figure \ref{fig:NEtiming} reports the results for Neuchâtel, another canton with a relatively late timing, which was on day 10 of the epidemic when the measures came into force. Concerning the effect of the lockdown timing on hospitalizations, we find a similar pattern as for Basel-Stadt. Albeit the effect on COVID-19-related fatalities is somewhat less pronounced, it turns statistically significant in the final periods of the evaluation window.

%\begin{figure}[htp]
%	\centering
%	\begin{subfigure}[b]{0.49\textwidth}
%		\includegraphics[width=\textwidth]{GRhosp.png}
%	\end{subfigure}
%	\begin{subfigure}[b]{0.49\textwidth}
%		\includegraphics[width=\textwidth]{GRdeath.png}
%	\end{subfigure}
%	\caption{{\small \textit{Effect of late timing of measures on cumulative hospitalizations (left) and deaths (right) per 10,000 inhabitants in Graubünden}}}\label{fig:GRtiming}
%\end{figure}
%Finally, Figure \ref{fig:GRtiming} reports the results for Graubünden, which was on day 10 of the epidemic and thus exposed to the measures four days earlier than in Basel-Stadt. Accordingly, we find somewhat less pronounced, but yet substantial effects on hospitalizations and fatilities that are statistically significant.

\subsection{Mobility Patterns in Switzerland} \label{mobility}

Table \ref{Tab:mobility} shows the average percentage change in distances traveled by MOBIS-COVID19 study participants  separately for 10 cantons during the COVID-19 pandemic when compared to a reference period of 4 weeks in autumn 2019, see \cite{Molloyetal2020}. While mobility was already reduced in times of comparably soft COVID-19 response measures before March 16th, the data point to a further and drastic reduction in mobility after the implementation of lockdown measures. The last column gives the mean difference before and after March 16th in terms of average percentage changes in distances traveled. On average, study participants reduced their mobility after March 16th by 57.17 \% when compared to the reference period in 2019 and, as indicated in the table, by 35.09 percentage points  more than in the first two weeks of March. A weighted regression of the pooled distance changes on an before-after-indicator for March 16th (with the weights corresponding to the canton-specific sample sizes) yields a heteroscedasticity-robust t-statistic of 10.85 for the before-after difference. The lockdown measures therefore had an immediate effect on mobility patterns that is highly statistically significant.

\begin{table}[!h]
	\begin{adjustwidth}{-1.8cm}{}
	{\footnotesize\begin{tabular}{l|c|cc||ccccccc|c}
		Canton & N   & Mar-02 & Mar-09 & Mar-16 & Mar-23 & Mar-30 & Apr-06 & Apr-13 & Apr-20 & Apr-27 & $\Delta_{Lockdown}$ \\
		\hline
		AR     & 55  & -25    & -41    & -71    & -58    & -50    & -57   & -55   &  -45   & -46   &  21.57 \\
		BL     & 142 & -15    & -11     & -62    & -61    & -60    & -61    & -56  & -54 & -50   &    44.71 \\
		BS     & 28  & -14   & -36    & -70    & -75    & -68    & -62   & -66   & -54 & -49   &   38.43 \\
		BE     & 145 & -31    & -36    & -67    & -60    & -57    & -57   & -51   & -48   & -43   &    21.21\\
		FR     & 6 & -61      & -23    & -63    & -56    & -61    & -52    & -44  & -65  & -54   &    14.43 \\
		GE     & 96 & 10      & -44    & -68    & -62    & -59    & -65   & -56   & -43  & -38   &    38.86 \\
		SZ     & 12  & -24     & -13     & -55    & -70    & -50    & -48   & -46   & -29   & -9   &    25.36 \\
		SO     & 14  & -13    & -41    & -62    & -65    & -53    & -49  & -50   &  -30   & -44   &   23.43 \\
		VD     & 228 & -8     & -22    & -65    & -70    & -68    & -65   & -64   & -55  & -55   &    48.14\\
		ZH     & 532 & -17    & -25    & -60    & -59    & -57    & -53   & -55   & -46  & -41   &   32.00 \\
		\hline
		\textbf{TOTAL}	& \textbf{1298} &        & \multicolumn{1}{l}{}   &  &   &   &  &  & &  & \textbf{ 35.09} \\
	\end{tabular}}
 \end{adjustwidth}
\caption{{\small \textit{Percentage change in distance traveled compared to reference period in 2019. Source:  \cite{Molloyetal2020}.}}}
\label{Tab:mobility}
\end{table}

\section{Conclusion}\label{conclusion}

In this paper, we analyzed the impact of lockdown measures on COVID-19 related fatalities and hospitalizations in Germany and Switzerland. For doing so, we exploited the fact that measures differed across regions and that the epidemic was more advanced in some regions than in others when certain measures came into force. Using OLS and doubly robust estimation, we compared the development of COVID-19-related hospitalization and death rates - two indicators which are arguably rather robust to regional differences in COVID-19 testing policies - across regions that have been at different epidemic stages when exposed to the lockdown measures.  For Switzerland, we also applied a synthetic control approach to investigate the impact of the relative timing of the lockdown in two selected cantons and investigated how the lockdown affected mobility patterns in a representative sample. In addition, we analyzed the impact of curfews as implemented in some German states on top of the federal ban on gatherings of more than 2 persons based on a cross-regional comparison.

For both countries, we found an earlier lockdown to be more effective than a later one, as cumulative hospitalisation and fatality rates measured relative to the region-specific start date of the epidemic were higher in regions with a more advanced spread of COVID-19 when the measures came into force. In contrast, our results did not provide evidence for curfews being more effective than bans on gatherings under the other lockdown measures already in place. Finally, we saw an immediate effect of the Swiss lockdown measures on behavioral patterns in terms of a significant reduction in mobility.

\pagebreak

\bibliographystyle{apacite}
\bibliography{Literatur}

\begin{thebibliography}{}

\bibitem [\protect \citeauthoryear {%
Abadie%
, Diamond%
\BCBL {}\ \BBA {} Hainmueller%
}{%
Abadie%
\ \protect \BOthers {.}}{%
{\protect \APACyear {2010}}%
}]{%
Abadieetal2010}
\APACinsertmetastar {%
Abadieetal2010}%
\begin{APACrefauthors}%
Abadie, A.%
, Diamond, A.%
\BCBL {}\ \BBA {} Hainmueller, J.%
\end{APACrefauthors}%
\unskip\
\newblock
\APACrefYearMonthDay{2010}{}{}.
\newblock
{\BBOQ}\APACrefatitle {Synthetic Control Methods for Comparative Case Studies:
  Estimating the Effect of California's Tobacco Control Program} {Synthetic
  control methods for comparative case studies: Estimating the effect of
  california's tobacco control program}.{\BBCQ}
\newblock
\APACjournalVolNumPages{Journal of the American Statistical
  Association}{105}{}{493-505,}.
\PrintBackRefs{\CurrentBib}

\bibitem [\protect \citeauthoryear {%
Abadie%
, Diamond%
\BCBL {}\ \BBA {} Hainmueller%
}{%
Abadie%
\ \protect \BOthers {.}}{%
{\protect \APACyear {2011}}%
}]{%
Abadieetal2011}
\APACinsertmetastar {%
Abadieetal2011}%
\begin{APACrefauthors}%
Abadie, A.%
, Diamond, A\BPBI J.%
\BCBL {}\ \BBA {} Hainmueller, J.%
\end{APACrefauthors}%
\unskip\
\newblock
\APACrefYearMonthDay{2011}{}{}.
\newblock
{\BBOQ}\APACrefatitle {Synth: An R Package for Synthetic Control Methods in
  Comparative Case Studies} {Synth: An r package for synthetic control methods
  in comparative case studies}.{\BBCQ}
\newblock
\APACjournalVolNumPages{Journal of Statistical Software}{42}{}{1-17}.
\PrintBackRefs{\CurrentBib}

\bibitem [\protect \citeauthoryear {%
Abadie%
\ \BBA {} Gardeazabal%
}{%
Abadie%
\ \BBA {} Gardeazabal%
}{%
{\protect \APACyear {2003}}%
}]{%
Abadie03}
\APACinsertmetastar {%
Abadie03}%
\begin{APACrefauthors}%
Abadie, A.%
\BCBT {}\ \BBA {} Gardeazabal, J.%
\end{APACrefauthors}%
\unskip\
\newblock
\APACrefYearMonthDay{2003}{}{}.
\newblock
{\BBOQ}\APACrefatitle {The Economic Costs of Conflict: A Case Study of the
  Basque Country} {The economic costs of conflict: A case study of the basque
  country}.{\BBCQ}
\newblock
\APACjournalVolNumPages{American Economic Review}{93}{}{1-22}.
\PrintBackRefs{\CurrentBib}

\bibitem [\protect \citeauthoryear {%
Amuedo-Dorantes%
, Borra%
, Garrido%
\BCBL {}\ \BBA {} Sevilla%
}{%
Amuedo-Dorantes%
\ \protect \BOthers {.}}{%
{\protect \APACyear {2020}}%
}]{%
amuedo2020timing}
\APACinsertmetastar {%
amuedo2020timing}%
\begin{APACrefauthors}%
Amuedo-Dorantes, C.%
, Borra, C.%
, Garrido, N\BPBI R.%
\BCBL {}\ \BBA {} Sevilla, A.%
\end{APACrefauthors}%
\unskip\
\newblock
\APACrefYearMonthDay{2020}{}{}.
\newblock
{\BBOQ}\APACrefatitle {Timing is Everything when Fighting a Pandemic: COVID-19
  Mortality in Spain} {Timing is everything when fighting a pandemic: Covid-19
  mortality in spain}.{\BBCQ}
\newblock
\APACjournalVolNumPages{IZA Discussion Paper Series}{}{}{}.
\PrintBackRefs{\CurrentBib}

\bibitem [\protect \citeauthoryear {%
Askitas%
, Tatsiramos%
\BCBL {}\ \BBA {} Verheyden%
}{%
Askitas%
\ \protect \BOthers {.}}{%
{\protect \APACyear {2020}}%
}]{%
AskitasTatsiramosVerheyden2020}
\APACinsertmetastar {%
AskitasTatsiramosVerheyden2020}%
\begin{APACrefauthors}%
Askitas, N.%
, Tatsiramos, K.%
\BCBL {}\ \BBA {} Verheyden, B.%
\end{APACrefauthors}%
\unskip\
\newblock
\APACrefYearMonthDay{2020}{}{}.
\newblock
{\BBOQ}\APACrefatitle {Lockdown Strategies, Mobility Patterns and COVID-19}
  {Lockdown strategies, mobility patterns and covid-19}.{\BBCQ}
\newblock
\APACjournalVolNumPages{IZA Discussion Paper No. 13293}{}{}{}.
\PrintBackRefs{\CurrentBib}

\bibitem [\protect \citeauthoryear {%
Banholzer%
\ \protect \BOthers {.}}{%
Banholzer%
\ \protect \BOthers {.}}{%
{\protect \APACyear {2020}}%
}]{%
Banholzer2020}
\APACinsertmetastar {%
Banholzer2020}%
\begin{APACrefauthors}%
Banholzer, N.%
, van Weenen, E.%
, Kratzwald, B.%
, Seeliger, A.%
, Tschernutter, D.%
, Bottrighi, P.%
\BDBL {}Vach, W.%
\end{APACrefauthors}%
\unskip\
\newblock
\APACrefYearMonthDay{2020}{}{}.
\newblock
{\BBOQ}\APACrefatitle {Estimating the impact of non-pharmaceutical
  interventions on documented infections with COVID-19: A cross-country
  analysis} {Estimating the impact of non-pharmaceutical interventions on
  documented infections with covid-19: A cross-country analysis}.{\BBCQ}
\newblock
\APACjournalVolNumPages{medRxiv}{}{}{}.
\PrintBackRefs{\CurrentBib}

\bibitem [\protect \citeauthoryear {%
Bayer%
\ \BBA {} Kuhn%
}{%
Bayer%
\ \BBA {} Kuhn%
}{%
{\protect \APACyear {2020}}%
}]{%
Bayer}
\APACinsertmetastar {%
Bayer}%
\begin{APACrefauthors}%
Bayer, C.%
\BCBT {}\ \BBA {} Kuhn, M.%
\end{APACrefauthors}%
\unskip\
\newblock
\APACrefYearMonthDay{2020}{April}{}.
\newblock
{\BBOQ}\APACrefatitle {Intergenerational Ties and Case Fatality Rates: A
  Cross-Country Analysis} {Intergenerational ties and case fatality rates: A
  cross-country analysis}.{\BBCQ}
\newblock
\APACjournalVolNumPages{IZA Discussion Paper Series}{}{}{}.
\PrintBackRefs{\CurrentBib}

\bibitem [\protect \citeauthoryear {%
Bicher%
, Rippinger%
, Urach%
, Brunmeir%
\BCBL {}\ \BBA {} Popper%
}{%
Bicher%
\ \protect \BOthers {.}}{%
{\protect \APACyear {2020}}%
}]{%
Bicheretal}
\APACinsertmetastar {%
Bicheretal}%
\begin{APACrefauthors}%
Bicher, M\BPBI R.%
, Rippinger, C.%
, Urach, C.%
, Brunmeir, D.%
\BCBL {}\ \BBA {} Popper, N.%
\end{APACrefauthors}%
\unskip\
\newblock
\APACrefYearMonthDay{2020}{}{}.
\newblock
{\BBOQ}\APACrefatitle {Agent-Based Simulation for Evaluation of Contact-Tracing
  Policies Against the Spread of SARS-CoV-2} {Agent-based simulation for
  evaluation of contact-tracing policies against the spread of
  sars-cov-2}.{\BBCQ}
\newblock
\APACjournalVolNumPages{medRxiv.}{}{}{}.
\PrintBackRefs{\CurrentBib}

\bibitem [\protect \citeauthoryear {%
Bonardi%
, Gallea%
, Kalanoski%
\BCBL {}\ \BBA {} Lalive%
}{%
Bonardi%
\ \protect \BOthers {.}}{%
{\protect \APACyear {2020}}%
}]{%
Bonardietal2020}
\APACinsertmetastar {%
Bonardietal2020}%
\begin{APACrefauthors}%
Bonardi, J\BHBI P.%
, Gallea, Q.%
, Kalanoski, D.%
\BCBL {}\ \BBA {} Lalive, R.%
\end{APACrefauthors}%
\unskip\
\newblock
\APACrefYearMonthDay{2020}{}{}.
\newblock
{\BBOQ}\APACrefatitle {Fast and local: How did lockdown policies affect the
  spread and severity of the COVID-19?} {Fast and local: How did lockdown
  policies affect the spread and severity of the covid-19?}{\BBCQ}
\newblock
\APACjournalVolNumPages{working paper, University of Lausanne}{}{}{}.
\PrintBackRefs{\CurrentBib}

\bibitem [\protect \citeauthoryear {%
Chernozhukov%
, Kasaha%
\BCBL {}\ \BBA {} Schrimpf%
}{%
Chernozhukov%
\ \protect \BOthers {.}}{%
{\protect \APACyear {2020}}%
}]{%
Chernozhukov2020}
\APACinsertmetastar {%
Chernozhukov2020}%
\begin{APACrefauthors}%
Chernozhukov, V.%
, Kasaha, H.%
\BCBL {}\ \BBA {} Schrimpf, P.%
\end{APACrefauthors}%
\unskip\
\newblock
\APACrefYearMonthDay{2020}{}{}.
\newblock
{\BBOQ}\APACrefatitle {Causal impact of masks, policies, behavior on early
  COVID-19 pandemic in the US} {Causal impact of masks, policies, behavior on
  early covid-19 pandemic in the us}.{\BBCQ}
\newblock
\APACjournalVolNumPages{medRxiv}{}{}{}.
\PrintBackRefs{\CurrentBib}

\bibitem [\protect \citeauthoryear {%
Dave%
, Friedson%
, Matsuzawa%
\BCBL {}\ \BBA {} Sabia%
}{%
Dave%
, Friedson%
, Matsuzawa%
\BCBL {}\ \BBA {} Sabia%
}{%
{\protect \APACyear {2020}}%
}]{%
Daveetal2020}
\APACinsertmetastar {%
Daveetal2020}%
\begin{APACrefauthors}%
Dave, D.%
, Friedson, A\BPBI I.%
, Matsuzawa, K.%
\BCBL {}\ \BBA {} Sabia, J\BPBI J.%
\end{APACrefauthors}%
\unskip\
\newblock
\APACrefYearMonthDay{2020}{}{}.
\newblock
{\BBOQ}\APACrefatitle {When Do Shelter-In-Place Orders Fight COVID-19 Best?
  Policy Heterogeneity across States and Adoption Time} {When do
  shelter-in-place orders fight covid-19 best? policy heterogeneity across
  states and adoption time}.{\BBCQ}
\newblock
\APACjournalVolNumPages{IZA Discussion Paper No. 13190}{}{}{}.
\PrintBackRefs{\CurrentBib}

\bibitem [\protect \citeauthoryear {%
Dave%
, Friedson%
, Matsuzawa%
, Sabia%
\BCBL {}\ \BBA {} Safford%
}{%
Dave%
, Friedson%
, Matsuzawa%
, Sabia%
\BCBL {}\ \BBA {} Safford%
}{%
{\protect \APACyear {2020}}%
}]{%
Daveetal2020b}
\APACinsertmetastar {%
Daveetal2020b}%
\begin{APACrefauthors}%
Dave, D.%
, Friedson, A\BPBI I.%
, Matsuzawa, K.%
, Sabia, J\BPBI J.%
\BCBL {}\ \BBA {} Safford, S.%
\end{APACrefauthors}%
\unskip\
\newblock
\APACrefYearMonthDay{2020}{}{}.
\newblock
{\BBOQ}\APACrefatitle {Were Urban Cowboys Enough to Control COVID-19? Local
  Shelter-in-Place Orders and Coronavirus Case Growth} {Were urban cowboys
  enough to control covid-19? local shelter-in-place orders and coronavirus
  case growth}.{\BBCQ}
\newblock

\PrintBackRefs{\CurrentBib}

\bibitem [\protect \citeauthoryear {%
Deb%
, Furceri%
, Ostry%
\BCBL {}\ \BBA {} Tawk%
}{%
Deb%
\ \protect \BOthers {.}}{%
{\protect \APACyear {2020}}%
}]{%
Debetal2020}
\APACinsertmetastar {%
Debetal2020}%
\begin{APACrefauthors}%
Deb, P.%
, Furceri, D.%
, Ostry, J\BPBI D.%
\BCBL {}\ \BBA {} Tawk, N.%
\end{APACrefauthors}%
\unskip\
\newblock
\APACrefYearMonthDay{2020}{May}{}.
\newblock
{\BBOQ}\APACrefatitle {The effect of containment measures on the COVID-19
  pandemic} {The effect of containment measures on the covid-19
  pandemic}.{\BBCQ}
\newblock
\APACjournalVolNumPages{CEPR Covid Economics, Vetted and Real-Time
  Papers}{19}{}{}.
\PrintBackRefs{\CurrentBib}

\bibitem [\protect \citeauthoryear {%
Donsimoni%
, Glawion%
, Plachter%
, Weiser%
\BCBL {}\ \BBA {} W\"{a}lde%
}{%
Donsimoni%
\ \protect \BOthers {.}}{%
{\protect \APACyear {2020}}%
}]{%
DonsimoniApr}
\APACinsertmetastar {%
DonsimoniApr}%
\begin{APACrefauthors}%
Donsimoni, J\BPBI R.%
, Glawion, R.%
, Plachter, B.%
, Weiser, C.%
\BCBL {}\ \BBA {} W\"{a}lde, K.%
\end{APACrefauthors}%
\unskip\
\newblock
\APACrefYearMonthDay{2020}{April}{}.
\newblock
{\BBOQ}\APACrefatitle {Should Contact Bans Be Lifted in Germany? A Quantitative
  Prediction of Its Effects} {Should contact bans be lifted in germany? a
  quantitative prediction of its effects}.{\BBCQ}
\newblock
\APACjournalVolNumPages{IZA Discussion Paper Series}{}{}{}.
\PrintBackRefs{\CurrentBib}

\bibitem [\protect \citeauthoryear {%
Fowler%
, Hill%
, Obradovich%
\BCBL {}\ \BBA {} Levin%
}{%
Fowler%
\ \protect \BOthers {.}}{%
{\protect \APACyear {2020}}%
}]{%
Fowleretal2020}
\APACinsertmetastar {%
Fowleretal2020}%
\begin{APACrefauthors}%
Fowler, J\BPBI H.%
, Hill, S\BPBI J.%
, Obradovich, N.%
\BCBL {}\ \BBA {} Levin, R.%
\end{APACrefauthors}%
\unskip\
\newblock
\APACrefYearMonthDay{2020}{}{}.
\newblock
{\BBOQ}\APACrefatitle {The Effect of Stay-at-Home Orders on COVID-19 Cases and
  Fatalities in the United States} {The effect of stay-at-home orders on
  covid-19 cases and fatalities in the united states}.{\BBCQ}
\newblock
\APACjournalVolNumPages{medRxiv preprint}{}{}{}.
\PrintBackRefs{\CurrentBib}

\bibitem [\protect \citeauthoryear {%
Friedson%
, McNichols%
, Sabia%
\BCBL {}\ \BBA {} Dave%
}{%
Friedson%
\ \protect \BOthers {.}}{%
{\protect \APACyear {2020}}%
}]{%
Friedsonetal2020}
\APACinsertmetastar {%
Friedsonetal2020}%
\begin{APACrefauthors}%
Friedson, A.%
, McNichols, D.%
, Sabia, J\BPBI J.%
\BCBL {}\ \BBA {} Dave, D.%
\end{APACrefauthors}%
\unskip\
\newblock
\APACrefYearMonthDay{2020}{}{}.
\newblock
{\BBOQ}\APACrefatitle {Did California's Shelter-in-Place Order Work? Early
  Coronavirus-Related Public Health Effects.} {Did california's
  shelter-in-place order work? early coronavirus-related public health
  effects.}{\BBCQ}
\newblock
\APACjournalVolNumPages{NBER Working Paper No. 26992}{}{}{}.
\PrintBackRefs{\CurrentBib}

\bibitem [\protect \citeauthoryear {%
Imbens%
\ \BBA {} Wooldridge%
}{%
Imbens%
\ \BBA {} Wooldridge%
}{%
{\protect \APACyear {2009}}%
}]{%
ImWo08}
\APACinsertmetastar {%
ImWo08}%
\begin{APACrefauthors}%
Imbens, G\BPBI W.%
\BCBT {}\ \BBA {} Wooldridge, J\BPBI M.%
\end{APACrefauthors}%
\unskip\
\newblock
\APACrefYearMonthDay{2009}{}{}.
\newblock
{\BBOQ}\APACrefatitle {Recent Developments in the Econometrics of Program
  Evaluation} {Recent developments in the econometrics of program
  evaluation}.{\BBCQ}
\newblock
\APACjournalVolNumPages{Journal of Economic Literature}{47}{}{5-86}.
\PrintBackRefs{\CurrentBib}

\bibitem [\protect \citeauthoryear {%
Juranek%
\ \BBA {} Zoutman%
}{%
Juranek%
\ \BBA {} Zoutman%
}{%
{\protect \APACyear {2020}}%
}]{%
JuranekZoutman2020}
\APACinsertmetastar {%
JuranekZoutman2020}%
\begin{APACrefauthors}%
Juranek, S.%
\BCBT {}\ \BBA {} Zoutman, F\BPBI T.%
\end{APACrefauthors}%
\unskip\
\newblock
\APACrefYearMonthDay{2020}{}{}.
\newblock
{\BBOQ}\APACrefatitle {The Effect of Social Distancing Measures on Intensive
  Care Occupancy: Evidence on COVID-19 in Scandinavia} {The effect of social
  distancing measures on intensive care occupancy: Evidence on covid-19 in
  scandinavia}.{\BBCQ}
\newblock
\APACjournalVolNumPages{FOR Discussion Paper 2/20, NHH Norwegion School of
  Economics}{}{}{}.
\PrintBackRefs{\CurrentBib}

\bibitem [\protect \citeauthoryear {%
Koo%
\ \protect \BOthers {.}}{%
Koo%
\ \protect \BOthers {.}}{%
{\protect \APACyear {2020}}%
}]{%
Singapore}
\APACinsertmetastar {%
Singapore}%
\begin{APACrefauthors}%
Koo%
, Cook%
, Park%
, Sun%
, Sun%
, Lim%
\BDBL {}Dickens%
\end{APACrefauthors}%
\unskip\
\newblock
\APACrefYearMonthDay{2020}{March}{}.
\newblock
{\BBOQ}\APACrefatitle {Interventions to mitigate early spread of SARS-CoV-2 in
  Singapore: a modelling study} {Interventions to mitigate early spread of
  sars-cov-2 in singapore: a modelling study}.{\BBCQ}
\newblock
\BIn{} Elsevier\ (\BED), \APACrefbtitle {The Lancet Infectious Diseases.} {The
  lancet infectious diseases.}
\newblock
\begin{APACrefURL} \url{https://doi.org/10.1016/S1473-3099(20)30162-6}
  \end{APACrefURL}
\PrintBackRefs{\CurrentBib}

\bibitem [\protect \citeauthoryear {%
Mitze%
, Kosfeld%
, Rode%
\BCBL {}\ \BBA {} W{\"a}lde%
}{%
Mitze%
\ \protect \BOthers {.}}{%
{\protect \APACyear {2020}}%
}]{%
Mitze2020}
\APACinsertmetastar {%
Mitze2020}%
\begin{APACrefauthors}%
Mitze, T.%
, Kosfeld, R.%
, Rode, J.%
\BCBL {}\ \BBA {} W{\"a}lde, K.%
\end{APACrefauthors}%
\unskip\
\newblock
\APACrefYearMonthDay{2020}{}{}.
\newblock
{\BBOQ}\APACrefatitle {Face Masks Considerably Reduce COVID-19 Cases in
  Germany: A Synthetic Control Method Approach} {Face masks considerably reduce
  covid-19 cases in germany: A synthetic control method approach}.{\BBCQ}
\newblock

\PrintBackRefs{\CurrentBib}

\bibitem [\protect \citeauthoryear {%
Molloy%
\ \protect \BOthers {.}}{%
Molloy%
\ \protect \BOthers {.}}{%
{\protect \APACyear {2020}}%
}]{%
Molloyetal2020}
\APACinsertmetastar {%
Molloyetal2020}%
\begin{APACrefauthors}%
Molloy, J.%
, Tchervenkov, C.%
, Schatzmann, T.%
, Schoeman, B.%
, Hintermann, B.%
\BCBL {}\ \BBA {} Axhausen, K\BPBI W.%
\end{APACrefauthors}%
\unskip\
\newblock
\APACrefYearMonthDay{2020}{}{}.
\newblock
{\BBOQ}\APACrefatitle {MOBIS-COVID19/05. Results as of 04/05/2020}
  {Mobis-covid19/05. results as of 04/05/2020}.{\BBCQ}
\newblock
\APACjournalVolNumPages{Arbeitsberichte Verkehrs- und Raumplanung 1498, ETH
  Zurich}{}{}{}.
\PrintBackRefs{\CurrentBib}

\bibitem [\protect \citeauthoryear {%
Qiu%
, Chen%
\BCBL {}\ \BBA {} Shi%
}{%
Qiu%
\ \protect \BOthers {.}}{%
{\protect \APACyear {2020}}%
}]{%
Qiu}
\APACinsertmetastar {%
Qiu}%
\begin{APACrefauthors}%
Qiu, Y.%
, Chen, X.%
\BCBL {}\ \BBA {} Shi, W.%
\end{APACrefauthors}%
\unskip\
\newblock
\APACrefYearMonthDay{2020}{April}{}.
\newblock
{\BBOQ}\APACrefatitle {Impacts of Social and Economic Factors on the
  Transmission of Coronavirus Disease 2019 (COVID-19) in China} {Impacts of
  social and economic factors on the transmission of coronavirus disease 2019
  (covid-19) in china}.{\BBCQ}
\newblock
\APACjournalVolNumPages{IZA Discussion Paper Series}{}{}{}.
\PrintBackRefs{\CurrentBib}

\bibitem [\protect \citeauthoryear {%
Robins%
\ \BBA {} Rotnitzky%
}{%
Robins%
\ \BBA {} Rotnitzky%
}{%
{\protect \APACyear {1995}}%
}]{%
RoRo95}
\APACinsertmetastar {%
RoRo95}%
\begin{APACrefauthors}%
Robins, J\BPBI M.%
\BCBT {}\ \BBA {} Rotnitzky, A.%
\end{APACrefauthors}%
\unskip\
\newblock
\APACrefYearMonthDay{1995}{}{}.
\newblock
{\BBOQ}\APACrefatitle {Semiparametric Efficiency in Multivariate Regression
  Models with Missing Data} {Semiparametric efficiency in multivariate
  regression models with missing data}.{\BBCQ}
\newblock
\APACjournalVolNumPages{Journal of the American Statistical
  Association}{90}{}{122-129}.
\PrintBackRefs{\CurrentBib}

\bibitem [\protect \citeauthoryear {%
Robins%
, Rotnitzky%
\BCBL {}\ \BBA {} Zhao%
}{%
Robins%
\ \protect \BOthers {.}}{%
{\protect \APACyear {1994}}%
}]{%
Robins+94}
\APACinsertmetastar {%
Robins+94}%
\begin{APACrefauthors}%
Robins, J\BPBI M.%
, Rotnitzky, A.%
\BCBL {}\ \BBA {} Zhao, L.%
\end{APACrefauthors}%
\unskip\
\newblock
\APACrefYearMonthDay{1994}{}{}.
\newblock
{\BBOQ}\APACrefatitle {Estimation of Regression Coefficients When Some
  Regressors Are not Always Observed} {Estimation of regression coefficients
  when some regressors are not always observed}.{\BBCQ}
\newblock
\APACjournalVolNumPages{Journal of the American Statistical
  Association}{90}{}{846-866}.
\PrintBackRefs{\CurrentBib}

\bibitem [\protect \citeauthoryear {%
Weber%
}{%
Weber%
}{%
{\protect \APACyear {2020}}%
}]{%
Weber2020}
\APACinsertmetastar {%
Weber2020}%
\begin{APACrefauthors}%
Weber, E.%
\end{APACrefauthors}%
\unskip\
\newblock
\APACrefYearMonthDay{2020}{June}{}.
\newblock
{\BBOQ}\APACrefatitle {Which measures flattened the curve in Germany?} {Which
  measures flattened the curve in germany?}{\BBCQ}
\newblock
\APACjournalVolNumPages{CEPR Covid Economics, Vetted and Real-Time
  Papers}{24}{}{}.
\PrintBackRefs{\CurrentBib}

\bibitem [\protect \citeauthoryear {%
Zetterqvist%
\ \BBA {} Sj{\"o}lander%
}{%
Zetterqvist%
\ \BBA {} Sj{\"o}lander%
}{%
{\protect \APACyear {2015}}%
}]{%
DoublyRobustEstimationwiththeRPackagedrgee}
\APACinsertmetastar {%
DoublyRobustEstimationwiththeRPackagedrgee}%
\begin{APACrefauthors}%
Zetterqvist, J.%
\BCBT {}\ \BBA {} Sj{\"o}lander, A.%
\end{APACrefauthors}%
\unskip\
\newblock
\APACrefYearMonthDay{2015}{}{}.
\newblock
{\BBOQ}\APACrefatitle {Doubly Robust Estimation with the R Package drgee}
  {Doubly robust estimation with the r package drgee}.{\BBCQ}
\newblock
\APACjournalVolNumPages{Epidemiologic Methods}{4}{}{}.
\PrintBackRefs{\CurrentBib}

\end{thebibliography}

\pagebreak

\renewcommand\appendix{\par
	\setcounter{section}{0}%
	\setcounter{table}{0}%
	\setcounter{figure}{0}%
	\renewcommand\thesection{\Alph{section}}%
	\renewcommand\thetable{\Alph{section}.\arabic{table}}}
\renewcommand\thefigure{\Alph{section}.\arabic{figure}}
\clearpage

\begin{appendix}
	
	\numberwithin{equation}{section}
	\noindent \textbf{\LARGE Appendices}

\section{Start Dates of Canton-Specific Epidemics} \label{app:Startdates}

\begin{table}[!h]
	\centering
		{\footnotesize\begin{tabular}{lc}
				\hline
				Canton & Start Date \\
				\hline
				Aargau (AG) & 03/16 \\
				Appenzell Innerrhoden (AI) & 03/13 \\
				Appenzell Ausserrhoden (AR) & 03/13 \\
				Bern (BE) & 03/14 \\
				Basel-Landschaft (BL) & 03/11  \\
				Basel-Stadt (BS) & 03/05 \\
				Fribourg (FR) & 03/11 \\
				Gen\`{e}ve (GE) & 03/09 \\
				Glarus (GL) & 03/12  \\
				Graub\"{u}nden (GR) & 03/09 \\
				Jura (JU) & 03/10 \\
				Luzern (LU) & 03/16  \\
				Neuch\^{a}tel (NE) & 03/07 \\
				Nidwalden (NW) & 03/09 \\
				Obwalden (OW) & 03/11 \\
				St. Gallen (SG) & 03/16 \\
				Schaffhausen (SH) & 03/17 \\
				Solothurn (SO) & 03/16 \\
				Schwyz (SZ) & 03/12 \\
				Thurgau (TG) & 03/16  \\
				Ticino (TI) & 03/05 \\
				Uri (UR) & 03/17  \\
				Vaud (VD) & 03/09 \\
				Valais (VS) & 03/12 \\
				Zug (ZG) & 03/13  \\
				Z\"{u}rich (ZH) & 03/12 \\
				Principality of Liechtenstein (LI) & 03/09 \\
				\hline
		\end{tabular}}

	\caption{{\small  \textit{2020 dates on which 1 confirmed infection per 10,000 inhabitants was reached in the Swiss cantons and LI.}}}
	\label{Tab:DCovariates}
\end{table}

\setcounter{table}{0}%
\setcounter{figure}{0}%
\clearpage

\section{Descriptive Statistics of Covariates} \label{app:Descriptives}

	\begin{table}[!h]
	\begin{adjustwidth}{-1.9cm}{}
		{\footnotesize\begin{tabular}{p{6cm}cccccc}
			\hline
			Variable & Total Sample & Late Timing & Intermediate Timing & Early Timing & Curfew & No Curfew\\
			& N = 408 & N = 81 & N = 275 & N = 52 & N = 149 & N = 259 \\
  \hline
Population & 203,103 & 276,529 & 197,295 & 119,444 & 158,786 & 228,598 \\
Population Density & 671 & 929 & 665 & 301 & 440 & 804 \\
Income per Capita (Euro) & 37,224 & 41,686 & 36,505 & 34,076 & 38,325 & 36,591 \\
Share of Population Aged 65+ & 0.222 & 0.208 & 0.221 & 0.244 & 0.226 & 0.219 \\
80+ Mortality Rate (per 1000 Inhabitants), 2017 & 6.52 & 5.96 & 6.52 & 7.36 & 6.68 & 6.42 \\
Share of Respiratory-Disease-Related Deaths, 2016 & 0.07 & 0.069 & 0.071 & 0.067 & 0.066 & 0.072 \\
Hospital Beds per 1000 Inhabitants & 6.31 & 6.08 & 6.25 & 6.97 & 6.69 & 6.09 \\
Share of Confirmed Infections Aged 80+ prior to Lockdown & 0.019 & 0.024 & 0.018 & 0.014 & 0.022 & 0.017 \\
Initial Growth Trend for Confirmed Cases in Log Points & 0.209 & 0.23 & 0.234 & 0.049 & 0.185 & 0.224 \\
Ban of events with $>$1000 Participants & 0.917 & 0.889 & 0.924 & 0.923 & 1 & 0.869 \\
Curfew & 0.365 & 0.247 & 0.378 & 0.481 & 1 & 0 \\
Ban of Groups of $>$5 Persons (prior to Contact Ban/Curfew) & 0.223 & 0.21 & 0.236 & 0.173 & 0 & 0.351 \\
Permission to Meet with 1 Non-Household-Member & 0.711 & 0.802 & 0.698 & 0.635 & 0.262 & 0.969 \\
\hline
		\end{tabular}}
	\end{adjustwidth}
	\caption{{\small  \textit{Mean of covariates considered in the estimations using the German data in the total sample, the late intervention group, the intermediate intervention group and the early intervention group, respectively.}}}
	\label{Tab:DCovariates}
\end{table}

\bigskip
	
	\begin{table}[ht]
		\begin{adjustwidth}{-1.4cm}{}
		{\footnotesize\begin{tabular}{p{8.5cm}cccc}
			\hline
			Variable & Total Sample & Late Timing & Intermediate Timing & Early Timing \\
			 & N = 27 & N = 8 & N = 11 & N = 8 \\
			\hline
			Population & 315,648 & 286,649 & 268,466 & 409,524 \\
			Population Density & 503 & 1,046 & 278 & 271 \\
			Income per Capita (CHF) & 80,404 & 102,840 & 73,134 & 67,964 \\
			Share of Population Aged 65+ & 0.192 & 0.193 & 0.19 & 0.193 \\
			Median Age of Confirmed Infections prior to Lockdown & 50.19 & 49.56 & 49.09 & 52.31 \\
			Initial Growth Trend of Confirmed Cases in Log Points & 0.235 & 0.239 & 0.21 & 0.266 \\
			Ban on Visits to Retirement Homes & 0.593 & 0.5 & 0.727 & 0.5 \\
			\hline
		\end{tabular}}
		\end{adjustwidth}
		\caption{{\small  \textit{Means of covariates considered in the estimations using the Swiss (and LI) data in the total sample, the late intervention group, the intermediate intervention group, the early intervention group, the group of counties with curfew and the group of counties without curfew respectively.}}}
		\label{Tab:Covariates}
	\end{table}

\setcounter{table}{0}%
\setcounter{figure}{0}%

\clearpage

\section{OLS Specifications for Germany and Switzerland} \label{app:OLSEstimates}

\begin{table}[!h]
	\centering
{\footnotesize\begin{tabular}{lcc}
	\hline
	& Estimate & Standard Error  \\
	\hline
	Intercept & -1.1628 & 0.6661  \\
	Intermediate Timing & 0.3348 & 0.1403  \\
	Late Timing & 0.5729 & 0.2663   \\
	Share of Population Aged 65+ & -6.4132 & 2.8281   \\
	Population: 0 - 105,878 & 0.4388 & 0.2112   \\
	Population: 105,879 - 158,080 & 0.2848 & 0.135   \\
	Population: 158,081 - 251,534 & 0.0665 & 0.0985   \\
	Population Density: 0 - 117.3 & 0.0801 & 0.1425   \\
	Population Density: 117.3 - 206.7 & 0.1201 & 0.1454   \\
	Population Density: 206.7 - 779.7 & 0.0613 & 0.1347   \\
	Income per Capita: 0 - 27,934 & -0.1437 & 0.1561   \\
	Income per Capita: 27,935 - 33,109 & -0.1721 & 0.1439   \\
	Income per Capita: 33,110 - 40,506 & 0.0568 & 0.1749   \\
	Share of Confirmed Infections Aged 80+ prior to Lockdown & 4.4466 & 2.1463   \\
	80+ Mortality Rate (per 1000 Inhabitants), 2017 & 0.2066 & 0.091   \\
	Share of Respiratory-Disease-Related Deaths, 2016 & 0.9538 & 3.7197   \\
	Hospital Beds per 1000 Inhabitants & -0.0329 & 0.0184   \\
	Initial Growth Trend for Confirmed Cases in Log Points: 0 - 0.14 & -0.1188 & 0.1852   \\
	Initial Growth Trend for Confirmed Cases in Log Points: 0.14 - 0.21 & -0.089 & 0.1407   \\
	Initial Growth Trend for Confirmed Cases in Log Points: 0.21 - 0.28 & -0.0369 & 0.136   \\
	Confirmed Infections per 10,000 Inhabitants on Epidemic Day 4 & 0.2556 & 0.0805   \\
	Recommendation against Events with $>$1000 Visitors & 0.1594 & 0.0985   \\
	Ban of Events with $>$1000 Visitors & 0.7132 & 0.141  \\
	Curfew & 0.2403 & 0.1111  \\
	\hline
\end{tabular}}
\caption{{\small  \textit{OLS estimates for Germany 28 days after the start of the county-specific epidemic with fatalities per 10,000 inhabitants as outcome variable.}}}
\label{Tab:DOLS}
\end{table}

\begin{table}[!h]
	\centering
	{\footnotesize\begin{tabular}{lcc}
		\hline
		& Estimate & Standard Error  \\
		\hline
Intercept & 39.2105 & 45.0916 \\
Intermediate Timing & 0.7961 & 0.7712 \\
Late Timing & 1.7187 & 0.6681 \\
Share of Population Aged 65+ & -337.2691 & 362.2737 \\
Squared Share of Population Aged 65+ & 848.3766 & 950.0775 \\
Population: 0 - 59,999 & -0.5647 & 1.1326 \\
Population Density & 4e-04 & 4e-04 \\
Income per Capita & 0 & 0 \\
Median Age of Confirmed Infections prior to Lockdown & -0.2783 & 1.308 \\
Squared Median Age of Confirmed Infections & 0.003 & 0.0131 \\
Initial Growth Trend for Confirmed Cases in Log Points & 6.3784 & 8.0649 \\
Confirmed Infections per 10,000 Inhabitants on Epidemic Day 4 & 0.0172 & 0.6938 \\
Ban on Visits to Retirement Homes & 0.153 & 0.4966 \\
\hline
	\end{tabular}}
	\caption{{\small  \textit{OLS estimates for Switzerland and LI 44 days after the start of the canton-specific epidemic with fatalities per 10,000 inhabitants as outcome variable.}}}
	\label{Tab:OLS}
\end{table}

\begin{table}[!h]
	\centering
	{\footnotesize\begin{tabular}{lcc}
			\hline
			& Estimate & Standard Error  \\
			\hline
 Intercept & -0.5481 & 0.5532 \\
 Curfew & 0.089 & 0.1081 \\
 Share of Population Aged 65+ & -5.6962 & 2.9762 \\
 Income per Capita: 0 - 27,934 & -0.0998 & 0.1872 \\
 Income per Capita: 27,935 - 33,109 & -0.0444 & 0.1598 \\
 Income per Capita: 33,110 - 40,506 & -0.056 & 0.1298 \\
 Population Density: 0 - 117.3 & 0.0077 & 0.1547 \\
 Population Density: 117.3 - 206.7 & 0.1532 & 0.1558 \\
 Population Density: 206.7 - 779.7 & 0.0388 & 0.1315 \\
 Population: 0 - 105,878 & 0.1964 & 0.1917 \\
 Population: 105,879 - 158,080 & 0.1198 & 0.1565 \\
 Population: 158,080 - 251,534 & -0.048 & 0.1067 \\
 Share of Confirmed Infections Aged 80+ & 0.6616 & 2.0497 \\
 80+ Mortality Rate (per 1000 Inhabitants), 2017 & 0.2029 & 0.0692 \\
 Share of Respiratory-Disease-Related Deaths, 2016 & 3.5314 & 3.6582 \\
 Hospital Beds per 1000 Inhabitants & -0.0201 & 0.016 \\
 Confirmed Fatalities per 10,000 Inhabitants 10 days before Curfew & -4.3731 & 4.1915 \\
 Confirmed Fatalities per 10,000 Inhabitants 5 days before Curfew & -2.7013 & 3.4591 \\
 Confirmed Fatalities per 10,000 Inhabitants 4 days before Curfew & 1.3937 & 3.7116 \\
 Confirmed Fatalities per 10,000 Inhabitants 3 days before Curfew & -2.8829 & 3.6353 \\
 Confirmed Fatalities per 10,000 Inhabitants 2 days before Curfew & 5.058 & 2.5642 \\
 Confirmed Fatalities per 10,000 Inhabitants 1 day before Curfew & 2.1268 & 2.1477 \\
 Confirmed Cases per 10,000 Inhabitants 25 days before Curfew & 2.478 & 4.2755 \\
 Confirmed Cases per 10,000 Inhabitants 20 days before Curfew & 0.9095 & 1.5009 \\
 Confirmed Cases per 10,000 Inhabitants 15 days before Curfew & 0.0804 & 0.4324 \\
 Confirmed Cases per 10,000 Inhabitants 10 days before Curfew & -0.3862 & 0.2614 \\
 Confirmed Cases per 10,000 Inhabitants 5 days before Curfew & 0.0339 & 0.2059 \\
 Confirmed Cases per 10,000 Inhabitants 4 days before Curfew & -0.3237 & 0.3682 \\
 Confirmed Cases per 10,000 Inhabitants 3 days before Curfew & 0.1382 & 0.3992 \\
 Confirmed Cases per 10,000 Inhabitants 2 days before Curfew & -0.148 & 0.2767 \\
 Confirmed Cases per 10,000 Inhabitants 1 day before Curfew & 0.3193 & 0.2064 \\
 Initial Growth Trend for Confirmed Cases in Log Points & 0.0158 & 0.0264 \\
 Recommendation against Events with $>$1000 Visitors & 0.2291 & 0.075 \\
 Ban of Events with $>$1000 Visitors & 0.6488 & 0.1937 \\
 Ban of Groups of >5 Persons (prior to Contact Ban/Curfew) & 0.1391 & 0.1265 \\
 Permission to Meet with 1 Non-Household-Member & -0.1802 & 0.1271 \\
	\hline
	\end{tabular}}
\caption{{\small  \textit{OLS estimates for the impact of curfews (compared to contact restrictions) 35 days after the imposition of curfews with fatalities per 10,000 inhabitants as outcome variable.}}}
\label{Tab:Ausgangssperre}
\end{table}

\setcounter{table}{0}%
\setcounter{figure}{0}%

\clearpage

\section{Estimations for Germany Without Covariates} \label{app:DOLS}

    \begin{figure}[h]
	\centering
	\begin{subfigure}[b]{0.49\textwidth}
		\includegraphics[width=\textwidth]{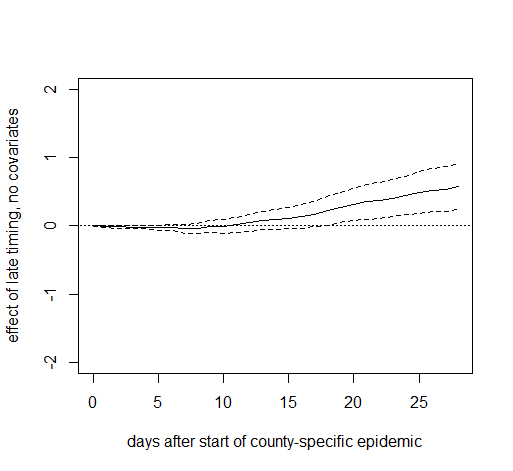}
	\end{subfigure}
	\begin{subfigure}[b]{0.49\textwidth}
		\includegraphics[width=\textwidth]{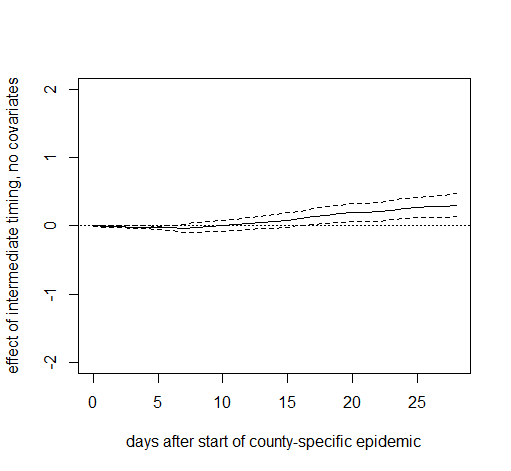}
	\end{subfigure}
	\caption{{\small \textit{OLS effects of late (left) and intermediate (right) timing of measures on cumulative deaths per 10,000 inhabitants without covariates.}}}\label{fig:Dtimingnocov}
	\end{figure}

\begin{figure}[!h]
	\centering
	\begin{subfigure}[b]{0.25\textwidth}
	\end{subfigure}
	\begin{subfigure}[b]{0.49\textwidth}
		\includegraphics[width=\textwidth]{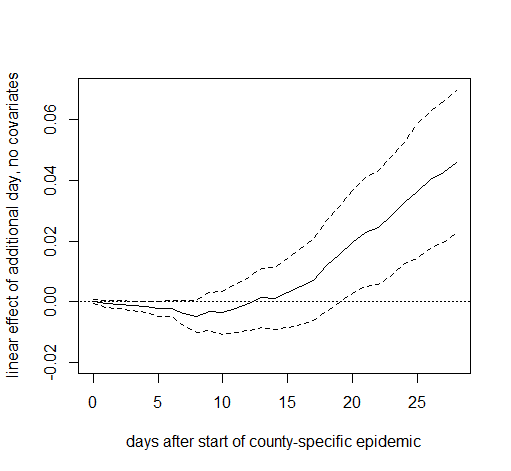}
	\end{subfigure}
	\begin{subfigure}[b]{0.25\textwidth}
	\end{subfigure}
	\caption{{\small \textit{OLS effect of delaying lockdown by one day on deaths per 10,000 inhabitants in Germany without covariates.}}}\label{fig:Dadddayapp}
\end{figure}

%\clearpage

%\section{Doubly Robust Semiparametric Estimations With Modified Group Definitions} \label{app:Dsemi}

\setcounter{table}{0}%
\setcounter{figure}{0}%

\clearpage
\section{Estimations for Switzerland Without Covariates and Including Ticino} \label{app:OLS}

\begin{figure}[h]
	\centering
	\begin{subfigure}[b]{0.49\textwidth}
		\includegraphics[width=\textwidth]{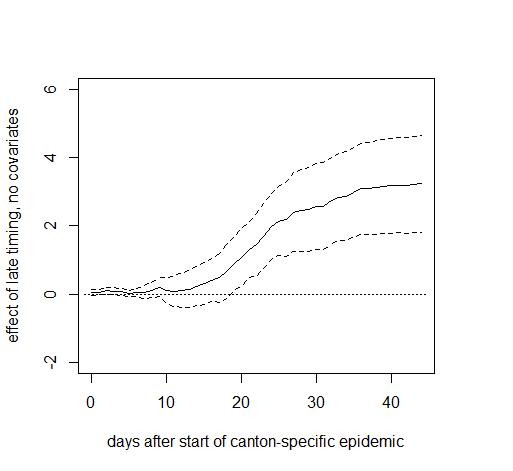}
	\end{subfigure}
	\begin{subfigure}[b]{0.49\textwidth}
		\includegraphics[width=\textwidth]{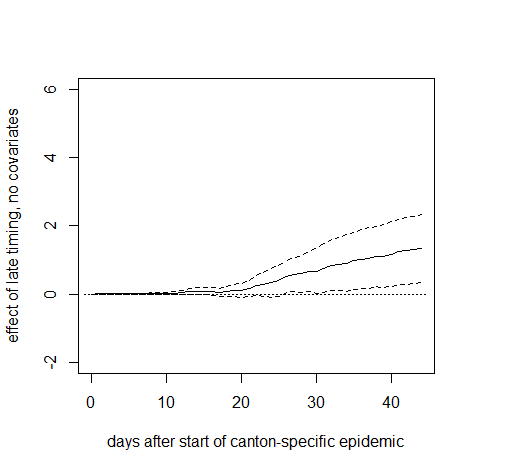}
	\end{subfigure}
	\caption{{\small \textit{OLS effect of late timing of measures on cumulative hospitalizations (left) and deaths (right) per 10,000 inhabitants without covariates excluding Ticino.}}}\label{fig:latetimingnocov}
\end{figure}

\begin{figure}[h]
	\centering
	\begin{subfigure}[b]{0.49\textwidth}
		\includegraphics[width=\textwidth]{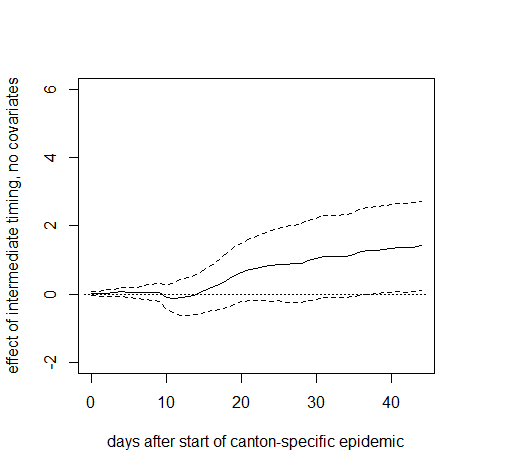}
	\end{subfigure}
	\begin{subfigure}[b]{0.49\textwidth}
		\includegraphics[width=\textwidth]{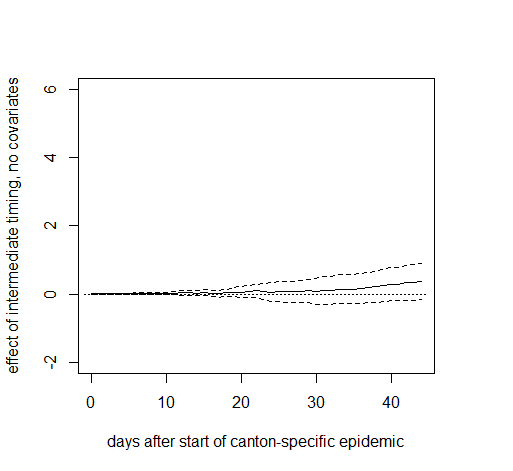}
	\end{subfigure}
	\caption{{\small \textit{OLS effect of intermediate timing of measures on cumulative hospitalizations (left) and deaths (right) per 10,000 inhabitants without covariates excluding Ticino.}}}\label{fig:intertimingnocov}
\end{figure}

\begin{figure}[h]
	\centering
	\begin{subfigure}[b]{0.49\textwidth}
		\includegraphics[width=\textwidth]{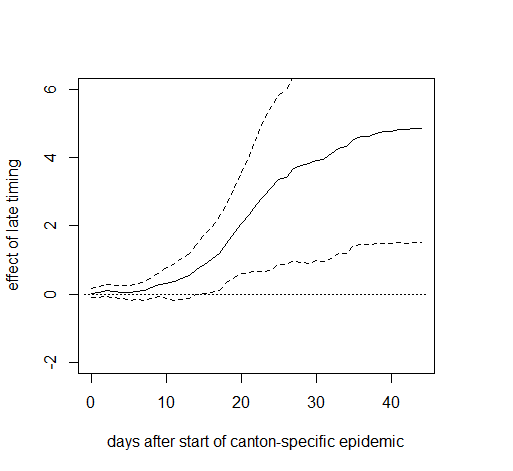}
	\end{subfigure}
	\begin{subfigure}[b]{0.49\textwidth}
		\includegraphics[width=\textwidth]{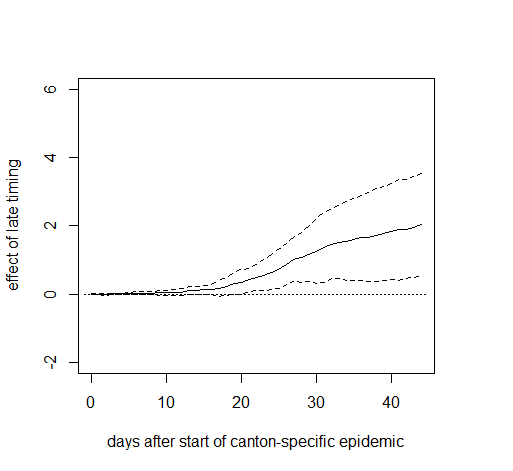}
	\end{subfigure}
	\caption{{\small \textit{OLS effect of late timing of measures on cumulative hospitalizations (left) and deaths (right) per 10,000 inhabitants with covariates including Ticino.}}}\label{fig:latetimingwithTI}
\end{figure}
\begin{figure}[!h]
	\centering
	\begin{subfigure}[b]{0.49\textwidth}
		\includegraphics[width=\textwidth]{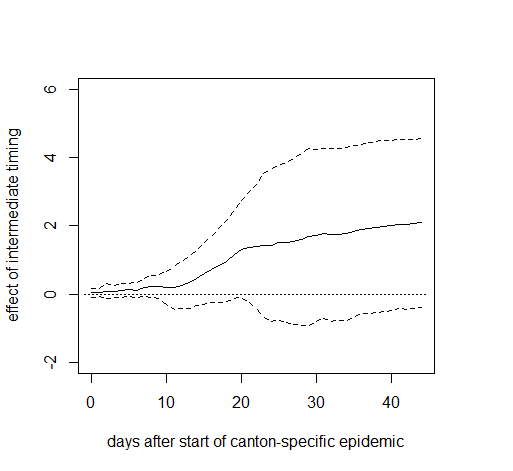}
	\end{subfigure}
	\begin{subfigure}[b]{0.49\textwidth}
		\includegraphics[width=\textwidth]{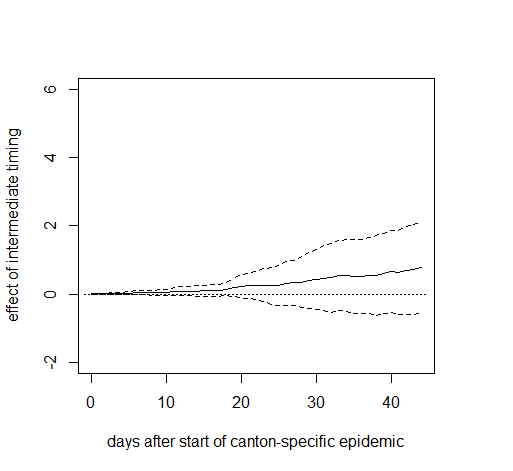}
	\end{subfigure}
	\caption{{\small \textit{OLS effect of intermediate timing of measures on cumulative hospitalizations (left) and deaths (right) per 10,000 inhabitants with covariates including Ticino.}}}\label{fig:intertimingwithTI}
\end{figure}

\end{appendix}
\end{document}